\documentclass[reprint,superscriptaddress,amsmath,amssymb,aps,prb]{revtex4-2}

\usepackage{xcolor}
\usepackage{graphicx}% Include figure files
\usepackage{dcolumn}% Align table columns on decimal point
\usepackage{bm}% bold math
\usepackage{xr}
\usepackage{lipsum}
\newcommand{\beginsupplement}{%
  \renewcommand{\thetable}{S\arabic{table}}%
  \renewcommand{\thefigure}{S\arabic{figure}}%
  \renewcommand{\thesection}{S\arabic{section}}%
  \renewcommand{\thesubsection}{S\arabic{section}.\arabic{subsection}}%   
}

\begin{document}

\title{Effect of Buried-Interface Preparation for Nb Superconducting Resonators on InP}

\author{Logan S. Kusher}
\affiliation{Center for Quantum Information Physics, New York University, New York, New York 10003, USA}

\author{Ding Peng}
\affiliation{Pacific Northwest National Laboratory, Richland, Washington 99354, USA}

\author{Zihua Zhu}
\affiliation{Pacific Northwest National Laboratory, Richland, Washington 99354, USA}

\author{Arunav Bordoloi}
\affiliation{Center for Quantum Information Physics, New York University, New York, New York 10003, USA}

\author{Axel Leblanc}
\affiliation{Center for Quantum Information Physics, New York University, New York, New York 10003, USA}

\author{Lukas J. Baker}
\affiliation{Center for Quantum Information Physics, New York University, New York, New York 10003, USA}

\author{Nichae Adnan}
\affiliation{Center for Quantum Information Physics, New York University, New York, New York 10003, USA}

\author{Jacob Issokson}
\affiliation{Center for Quantum Information Physics, New York University, New York, New York 10003, USA}

\author{Alvin Wang}
\affiliation{Center for Quantum Information Physics, New York University, New York, New York 10003, USA}

\author{Frederik Knudsen}
\affiliation{Center for Quantum Information Physics, New York University, New York, New York 10003, USA}

\author{Krishna Dindial}
\affiliation{Center for Quantum Information Physics, New York University, New York, New York 10003, USA}

\author{Melissa Mikalsen}
\affiliation{Center for Quantum Information Physics, New York University, New York, New York 10003, USA}

\author{Taha Kaleem}
\affiliation{Center for Quantum Information Physics, New York University, New York, New York 10003, USA}

\author{Andrei Vrajitoarea}
\affiliation{Center for Quantum Information Physics, New York University, New York, New York 10003, USA}

\author{Yingge Du}
\affiliation{Pacific Northwest National Laboratory, Richland, Washington 99354, USA}

\author{Patrick J. Strohbeen}
\thanks{Now at Research Laboratory of Electronics, MIT}
\affiliation{Center for Quantum Information Physics, New York University, New York, New York 10003, USA}

\author{Javad Shabani}
\thanks{Corresponding author: jshabani@nyu.edu}
\affiliation{Center for Quantum Information Physics, New York University, New York, New York 10003, USA}

\date{\today}

\begin{abstract}
Substrate surface preparation is a key step in the fabrication of electronic devices. In III--V semiconductor platforms, e.g. used in HEMTs and lasers, removal of the native substrate oxide is extremely important, where improper removal will negatively affect end-of-line device performance. However, the impact of substrate preparation in hybrid superconductor--semiconductor (S--Sm) systems relevant to quantum information applications remains poorly understood. This study compares three surface preparations for the deposition of sputtered Nb films on InP: (i) no intentional oxide removal (control), (ii) \textit{in-situ} $\mathrm{Ar}^{+}$ milling, and (iii) S-passivation. \textit{In-situ} $\mathrm{Ar}^{+}$ milling reduces the O concentration at the metal--substrate (MS) interface, but also roughens the InP surface, increasing the effective thickness of the Nb--InP interface and promoting O incorporation through extended defects in the Nb film. S-passivation suppresses interfacial O more effectively while preserving a sharper and smoother buried interface. It also yields Nb films with higher superconducting transition temperatures and less structural damage than the $\mathrm{Ar}^{+}$-milled samples. Despite these materials improvements, the microwave response is comparable across the three preparations. At single photon powers, the highest internal quality factors, $Q_i$, are approximately $1.30\times10^{5}$ ($130\mathrm{k}$) for the control resonators, $9.7\times10^{4}$ ($97\mathrm{k}$) for the S-passivated resonators, and $8.4\times10^{4}$ ($84\mathrm{k}$) for the $\mathrm{Ar}^{+}$-milled resonators. These results suggest that the present devices are not primarily limited by dielectric loss at the buried Nb--InP interface. Instead, aggressive oxide-removal processes can introduce additional damage and degrade performance, while the dominant microwave loss is likely associated with other channels such as substrate, quasiparticle, or package-related losses.
\end{abstract}

\maketitle

\section{Introduction}

Lossy interfacial regions are a major source of microwave dissipation in superconducting resonators and qubits~\cite{gao2008experimental,wang2015surface,woods2019determining,mcrae2020materials}. At millikelvin temperatures and single-photon powers, resonant absorption by parasitic two-level systems (TLSs) is a leading explanation for this loss. Although the microscopic origins of TLSs remain an active area of research, studies have associated them with amorphous and native oxides, adsorbed contaminants, lithographic residue, and process-modified interfacial layers~\cite{martinis2005decoherence,gao2008experimental,muller2019tls,bal2024encapsulation,verjauw2021oxide,quintana2014microfab,murthy2022tofsims,pearton1990ion}. In planar superconducting circuits, microwave modes of interest concentrate near metal edges and strongly overlap with nanometer-scale lossy layers, allowing their loss contribution to exceed that of the bulk materials at the metal--air (MA), substrate--air (SA), and metal--substrate (MS) interfaces~\cite{wenner2011surface,wang2015surface,woods2019determining,calusine2018analysis}.

Much of the previous work on interfacial microwave loss has focused on the exposed MA and SA interfaces. Trenching reduces the participation of the SA interface in high-field regions~\cite{bruno2015reducing,calusine2018analysis}. Surface passivation and encapsulation suppress native-oxide formation at exposed MA interfaces~\cite{bal2024encapsulation,chang2025noble,gupta2026nbpassivation}. Changes to the superconducting metal, most notably the use of Ta, can also improve coherence through changes in surface chemistry~\cite{place2021tantalum,chang2025noble}. By comparison, the buried MS interface is substantially more difficult to study and control. Once this interface is formed, it cannot be cleaned, capped, or chemically altered in the same manner as an exposed MA or SA interface. Its properties are instead determined by the substrate surface before deposition, the pre-deposition treatment, and the initial nucleation of the superconducting film.

This buried interface is particularly relevant to hybrid superconductor--semiconductor quantum devices. In planar Al--InAs heterostructures, the superconducting proximity effect creates a gate-tunable Josephson junction within a two-dimensional electron gas~\cite{shabani2016prb}. Qubits based on these voltage-controlled junctions, commonly referred to as gatemon qubits, have reported energy-relaxation times ranging from hundreds of nanoseconds to approximately $2~\mu\mathrm{s}$ in planar InAs 2DEG devices~\cite{casparis2018gatemon,strickland2024losses}. By comparison, Al--InAs nanowire gatemons have reached $T_1=5.3~\mu\mathrm{s}$~\cite{casparis2016gatemon}, while Sn--InAs nanowire transmons have reached $T_1=26.9\pm0.7~\mu\mathrm{s}$~\cite{purkayastha2025sn}.

The difference between the Al--InAs and Sn--InAs nanowire devices is consistent with substantial dissipation at or near the Al--InAs interface~\cite{casparis2016gatemon,purkayastha2025sn}. The further reduction in $T_1$ between Al--InAs nanowire and planar devices points to additional loss associated with the epitaxial heterostructure, including structural defects and dislocations~\cite{casparis2018gatemon,strickland2024losses,liu2026strongly}. Dissipation within the semiconductor weak link may provide another contribution~\cite{sun2026junction}.

The InP substrate introduces an additional platform-level loss channel because it is piezoelectric. Microwave electric fields can couple to acoustic modes and radiate energy into the substrate~\cite{scigliuzzo2020phononic,yang2023piezoelectric}. This channel is much weaker or absent in common low-loss substrates such as high-resistivity Si and sapphire~\cite{scigliuzzo2020phononic,yang2023piezoelectric}. Similar piezoelectric radiation loss has been observed in GaAs, where it can limit superconducting resonator quality factors~\cite{scigliuzzo2020phononic}. Piezoelectric coupling may therefore impose a substrate-level ceiling on devices fabricated on InP. This number is estimated to be 10$\mu$s.~\cite{scigliuzzo2020phononic} However, the substantially shorter $T_1$ values of planar gatemons suggest that current devices may instead be limited by additional loss from the buried interface, heterostructure, gates, or semiconductor weak link.

Because the full gatemon stack contains several possible sources of dissipation, it is difficult to isolate the dominant loss mechanism directly from qubit measurements. We therefore study a simplified superconductor–InP system that isolates the buried superconductor–substrate interface from the additional complexity of gates and semiconductor weak links. Previous measurements of Al-based resonators on InAs/InP heterostructures reported low-power internal quality factors up to approximately \(4\times10^4\)~\cite{strickland2024losses}. We choose Nb instead of Al because its larger superconducting gap makes it promising for future InAs/InP hybrid devices and because it is widely used in superconducting microwave circuits~\cite{bal2024encapsulation}.

\begin{figure}[htbp!]
    \centering
    \includegraphics[width=\linewidth]{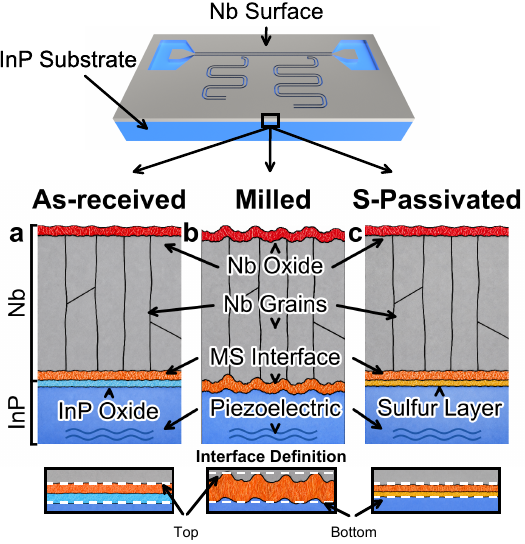}
    \caption{
    Microwave loss channels in CPW resonators fabricated from sputtered Nb on epitaxy-ready InP. A three-dimensional rendering of the CPW chip is shown above three cross-sectional schematics of the experimental thin-film structures described in panels (a)–(c). Nb grain boundaries, amorphous MS interfaces, native oxides, and piezoelectric losses in the InP substrate are highlighted as possible microwave loss channels. 
    (a) Sputtered Nb on InP, where the native InP surface oxide remains at the buried MS interface and contributes to an amorphous interfacial region. 
    (b) Ar$^+$-milled Nb on InP, where ion milling removes the native oxide but roughens the already amorphous MS interface.
    (c) S-passivated Nb on InP, where chemical treatment and subsequent passivation with a sulfur layer removes the native oxide and suppresses reoxidation while preserving a smoother buried interface than that produced by the Ar$^+$-milling process.
    }
    \label{fig:fig-one}
\end{figure}

Here, we investigate the role of the buried Nb--InP interface on the resulting Nb film microstructure, superconducting transport behavior, and microwave loss. To do this, we compare three substrate preparation methods, depicted in Figure~\ref{fig:fig-one}: as-received, Ar-milled, and S-passivated \cite{lee2019facile}. After substrate preparation, Nb films are deposited at room temperature utilizing DC magnetron sputtering (AJA International Inc.). In the milled case, the milling and deposition process occurs within the same high-vacuum sputtering chamber (Base Pressure $\sim 5\times10^{-9}~\textup{Torr}$) to maintain a clean MS interface. For the study presented here, ion milling was conducted at a dynamic Ar pressure of 3~mTorr with a plasma power of 25~W. S-passivated samples are prepared in an acid hood within a cleanroom environment located in the same building as the sputtering chamber. These substrates are first etched for 30~s in a dilute buffered oxide etchant prepared by mixing 10~mL of Transene BOE~6:1 stock solution with 70~mL of deionized water. They are then quickly transferred into an aqueous 10\% $(\mathrm{NH}_{4})_{2}\mathrm{S}$ solution (Thermo Fisher Scientific) to passivate the surface with S before Nb deposition. Time-of-flight secondary-ion mass spectrometry (ToF-SIMS) is used to measure chemical composition of the Nb film, the buried interface, and the underlying substrate. Grain structure as a function of substrate preparation is evaluated through \textit{ex-situ} atomic force microscopy (AFM). Cross-sectional scanning transmission electron microscopy (STEM), STEM energy-dispersive X-ray spectroscopy (STEM-EDS), and cluster-based 4D-STEM grain mapping are used to examine the buried interface, film structure, grain structure and amorphous regions, and O-rich regions. Lastly, these materials are fabricated into coplanar waveguide (CPW) resonators to evaluate the differences in microwave loss due to the substrate preparation method.

\section{\label{sec:level1}Results and Discussion}

\begin{figure*}[htbp!]
    \centering
    \includegraphics[width=7.0in]{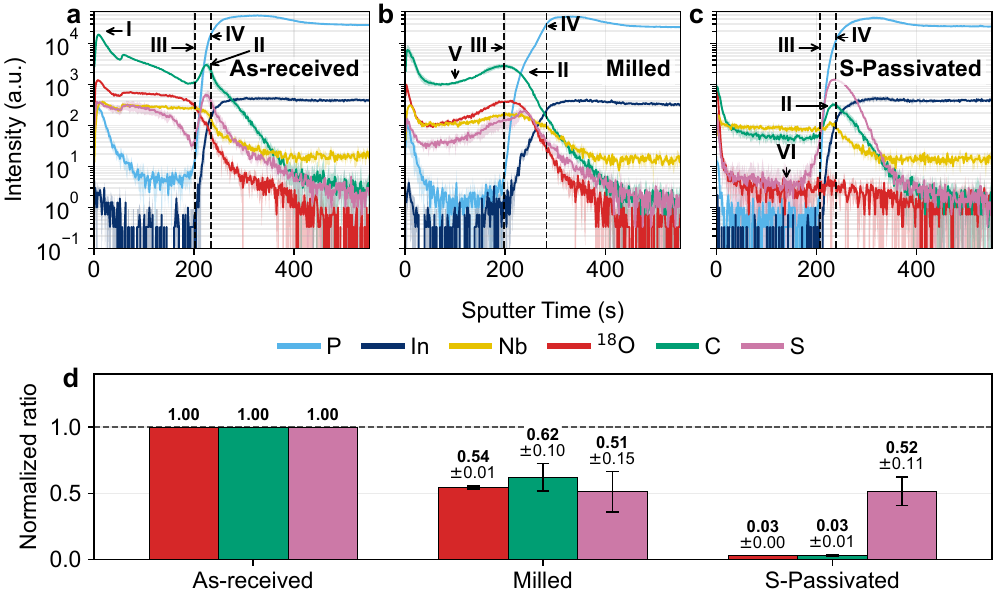}
    \caption{
    ToF-SIMS depth profiles and normalized impurity incorporation metrics for Nb-on-InP~(001) samples. Panels \textbf{(a)}--\textbf{(c)} show averaged depth profiles for the as-received, Ar-milled, and S-passivated samples, respectively. Data were acquired at four locations across each treatment condition and averaged, with $\pm 1$ standard deviation shown as the shaded region around each ion trace. The two vertical dashed lines in each depth-profile panel mark the bounds of the buried Nb--InP interface transition region, as defined using the piecewise-linear construction shown in Figure~\ref{fig:s_sims_interface_method}. The film-side and substrate-side bounds of the Nb--InP interface are indicated by markers III and IV, respectively, and their separation is used as a measure of apparent interface broadening in sputter time. Panel \textbf{(d)} summarizes the normalized integrated $^{18}$O:In, C:In, and S:In ratios for each surface preparation. Ratios were normalized to the as-received sample. The uncertainty values in panel \textbf{(d)} were calculated by first computing the integrated species:In ratio separately for each measured replicate trace and then taking the standard deviation of those replicate ratios after normalization to the as-received condition.  
    }
    \label{fig:sims_depth_profiles}
\end{figure*}

To evaluate how surface preparation changes contaminant incorporation in the Nb films, we use time-of-flight secondary-ion mass spectrometry (ToF-SIMS), which is well suited for identifying O and C species in Nb thin films~\cite{murthy2022tofsims,bose2020nbtofsims}. ToF-SIMS measurements were performed at the Environmental Molecular Sciences Laboratory (EMSL), located at Pacific Northwest National Laboratory. ToF-SIMS methods are described in Appendix~\ref{SIMS_Measurement}.

\begin{table}[htbp!]
    \centering
    \caption{
    Approximate interface widths extracted from the ToF-SIMS depth profiles using the In-derived Nb--InP interface bounds. The interface width was calculated as the separation between the upper and lower interface bounds, (III) and (IV). Approximate interface widths in nm were estimated by scaling the sputter-time axis using the AFM-measured Nb thickness of 130~nm, with the midpoint of the interface region taken as the bottom of the Nb film.
    }
    \label{sims_tab_in}
    \begin{tabular}{lccc}
        \hline
        & As-received & Milled & S-Passivated \\
        \hline
        Interface Width (s) & 32 & 85 & 31 \\
        Interface Width (nm) & 19 & 46 & 18 \\
        \hline
    \end{tabular}
\end{table}

\begin{figure*}[htbp!]
    \centering
    \includegraphics[width=7in]{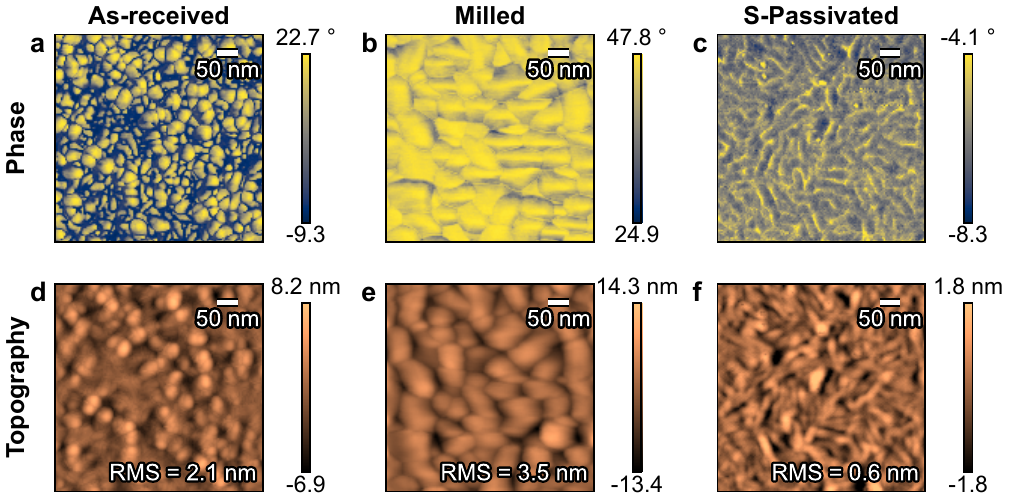}
    \caption{Atomic force microscopy of Nb films deposited on InP~(001) after different surface preparations. Panels (a)--(c) show top-surface phase images for the as-received, milled, and S-passivated samples, respectively, used for grain analysis. Panels (d)--(f) show the corresponding topography images for the as-received, milled, and S-passivated samples, respectively, with the RMS roughness values overlaid.}
    \label{fig:afm_phase_topography}
\end{figure*}

Figure~\ref{fig:sims_depth_profiles}(a)--(c) shows the ToF-SIMS depth profiles for the as-received, Ar-milled, and S-passivated samples. In each panel, marker (III) defines the film-side bound of the buried Nb--InP interface region, and marker (IV) defines the substrate-side bound where the In signal reaches the InP substrate plateau. The normalized integrated impurity metrics shown in Fig.~\ref{fig:sims_depth_profiles}(d) were calculated using these same dashed-line-defined regions. For each species $X = {}^{18}\mathrm{O}$, C, or S, the numerator was calculated as $\int_{t_0}^{t_{\mathrm{IV}}} I_X(t)\,dt$, where $t_0$ is the beginning of the sputter profile and $t_{\mathrm{IV}}$ is the lower bound of the Nb--InP interface region. This window includes the Nb surface, the Nb film, and the III--IV interface region. The In normalization term was calculated as $\int_{t_{\mathrm{IV}}}^{t_{\mathrm{IV}}+200~\mathrm{s}} I_{\mathrm{In}}(t)\,dt$, corresponding to the In signal from the InP substrate immediately below the interface. The species-to-In ratio was therefore defined as $R_X=\int_{t_0}^{t_{\mathrm{IV}}} I_X(t)\,dt/\int_{t_{\mathrm{IV}}}^{t_{\mathrm{IV}}+200~\mathrm{s}} I_{\mathrm{In}}(t)\,dt$. Each ratio was then normalized to the corresponding as-received value, $\widetilde{R}_X=R_X/R_{X,\mathrm{as-received}}$, so that the as-received sample defines a reference value of 1 for each species. For replicate traces, the uncertainty was calculated as the standard deviation of the individually integrated replicate ratios.

Across all three depth profiles, the uncapped Nb--air surface shows strong C and O peaks at (I). This is expected because the top Nb surface is a free surface that was not capped, so these surface peaks do not, by themselves, distinguish the preparation methods. Similar O- and C-containing species have been reported in ToF-SIMS studies of Nb films, where such signals were associated with surface oxides, hydrocarbon-related contamination, and process-dependent impurity incorporation~\cite{murthy2022tofsims,bose2020nbtofsims}. The main differences between the samples appear within the Nb film and near the buried Nb--InP interface. In the as-received sample, O remains present throughout the Nb film, and the integrated $^{18}\mathrm{O}$:In, C:In, and S:In ratios define the normalization reference values of 1.00 in Fig.~\ref{fig:sims_depth_profiles}(d). Ar milling reduces the normalized integrated $^{18}\mathrm{O}$:In and C:In ratios to $0.54 \pm 0.01$ and $0.62 \pm 0.10$, respectively, relative to the as-received sample. The C signal also forms a broadened feature near the interface region, as indicated by (II), and is smeared throughout the Nb film, as indicated by (V). The S-passivated sample shows the strongest reduction in both O and C incorporation, with normalized $^{18}\mathrm{O}$:In and C:In ratios of $0.03 \pm 0.00$ and $0.03 \pm 0.01$, respectively. This is consistent with the O intensity dropping to background levels throughout the Nb film and the absence of an O peak at the buried interface, as indicated by (VI). The normalized S:In ratios for the Ar-milled and S-passivated samples are similar, $0.51 \pm 0.15$ and $0.52 \pm 0.11$, respectively, so this depth-integrated S metric is treated as a relative sulfur-containing signal over the Nb-plus-interface region rather than as a direct measure of a localized interfacial sulfur layer.

\begin{figure*}[htbp!]
    \includegraphics[width=7in]{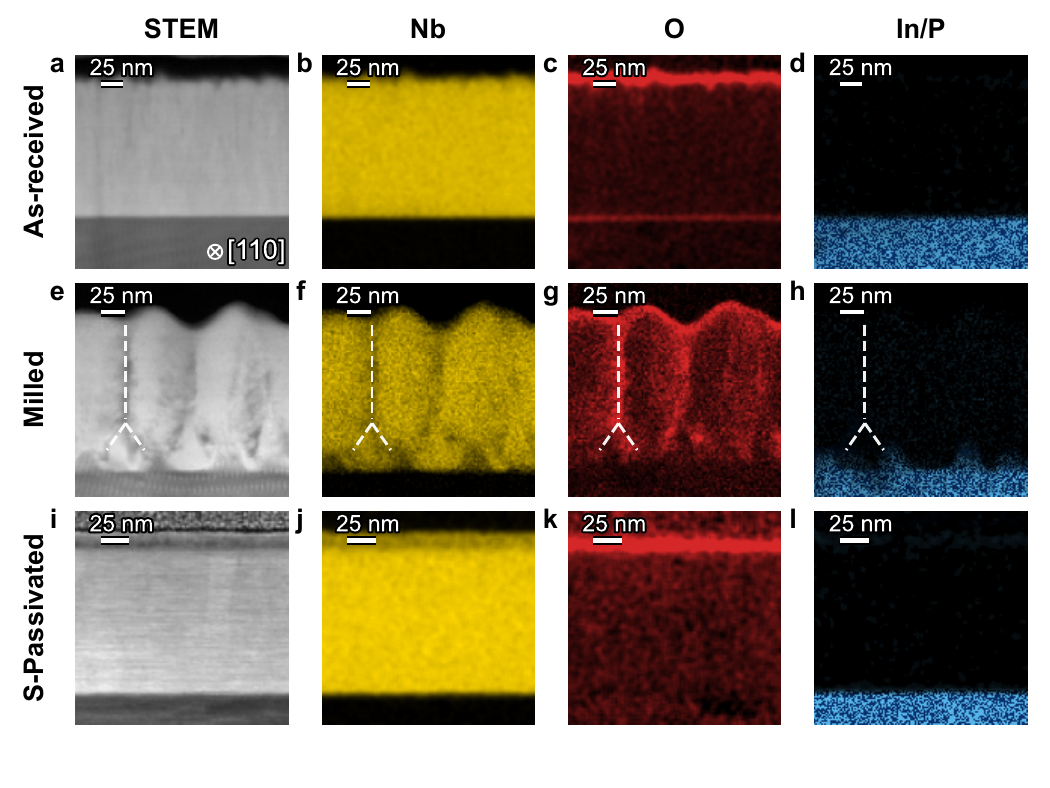}
    \caption{Cross-sectional HAADF-STEM and EDS comparison of the buried Nb/InP~(001) interface for the \textbf{(a)}--\textbf{(d)} as-received, \textbf{(e)}--\textbf{(h)} milled, and \textbf{(i)}--\textbf{(l)} S-passivated substrate treatments. \textbf{(a)}, \textbf{(e)}, and \textbf{(i)} present the HAADF-STEM images of the cross section for the as-received, milled, and S-passivated samples, respectively. EDS data for the Nb signal are presented in \textbf{(b)}, \textbf{(f)}, and \textbf{(j)}, the O signal in \textbf{(c)}, \textbf{(g)}, and \textbf{(k)}, and the combined In/P signal in \textbf{(d)}, \textbf{(h)}, and \textbf{(l)}. All images are taken in the [110] cross-section direction. The milled interface exhibits pronounced roughening of the InP surface, forming pyramidal faceted structures at the buried interface. This roughened morphology propagates into the deposited Nb film, producing a nonplanar interface, increased surface roughness, and clearly visible grain boundaries throughout the film thickness. The dashed vertical lines indicate the locations of InP pyramids. The O signal is preferentially localized between these vertical lines, suggesting enhanced O incorporation and segregation at boundary sites formed between adjacent pyramidal features.
    In contrast to the milled sample, the S-passivated interface remains sharp and planar, with no evidence of pyramidal InP surface faceting. The Nb film appears substantially more continuous, with no pronounced O-rich boundary features in the examined field of view, indicating that S-passivation preserves interface planarity and suppresses the microstructural degradation induced by argon milling.
}
    \label{fig:tem_8panel}
\end{figure*}

Interface bounds were independently extracted from the In and P depth profiles using the analyses described in Appendix~\ref{SIMS_Interfaces}. For each indicator, the same derived bounds were used to calculate both the contaminant ratios and the apparent interface widths. The following discussion uses the In-derived results because In is less volatile than P and provides a clearer interface transition. For the ratios normalized to Phosphorous intensity, see Table~\ref{s_P_sims_tab}. Ultimately, the ratios, regardless of the normalization species In or P, are qualitatively the same. The interface widths summarized in Table~\ref{sims_tab_in} show that the reduction in $^{18}\mathrm{O}$ incorporation in the Ar-milled sample is accompanied by substantial broadening of the buried Nb--InP interface. By subtracting the sputter time at the upper interface bound (III) from that at the lower interface bound (IV), using the In-derived dashed-line positions, the III--IV separation for the as-received sample is approximately 32~s, corresponding to an approximate interface width of 19~nm. After Ar milling, this region broadens to approximately 85~s, corresponding to an approximate width of 46~nm. This is approximately 2.7$\times$ the as-received interface width in sputter time, or approximately 2.4$\times$ in the calibrated length scale, indicating substantial broadening of the buried Nb--InP interface after Ar milling. The TEM image of the milled substrate in Figure~\ref{fig:tem_8panel}(h) shows approximately 25~nm pyramidal features extending from the nominal InP surface. When this roughness scale is combined with the as-received ToF-SIMS-estimated interface width of approximately 19~nm, the resulting length scale is approximately 44~nm, which is reasonably consistent with the 46~nm milled interface width estimated from the ToF-SIMS analysis. This interpretation is consistent with prior reports showing that Ar$^+$ ion milling and sputtering can modify InP through near-surface damage and sputter-induced roughness evolution~\cite{pearton1990ion,frost2000inp}. In contrast, the S-passivated sample has a III--IV separation of approximately 31~s, corresponding to an approximate width of 18~nm. This width is close to the as-received case and much narrower than the Ar-milled sample, consistent with a sharper buried Nb--InP interface with reduced broadening. The reduced $^{18}\mathrm{O}$ signal in the S-passivated sample is also consistent with prior studies showing that ammonium sulfide treatment of InP removes or suppresses native oxide formation and improves the chemical and electrical quality of the passivated InP surface~\cite{tian2014inp,lee2019facile}.

The AFM phase-contrast and topography data for the resulting Nb films on the three substrate preparations are shown in Figure~\ref{fig:afm_phase_topography}. The phase-contrast images reveal clear differences in Nb grain morphology across the as-received/control, Ar-milled, and S-passivated samples. The as-received/control film exhibits the smallest measured mean grain size, with a mean equivalent diameter of $17.4 \pm 8.2$~nm and the highest grain count of 439 grains within the scan window. In contrast, the Ar-milled film exhibits the largest characteristic grain size and the lowest grain count, with a mean equivalent diameter of $60.9 \pm 18.1$~nm across 57 grains, corresponding to an increase in grain size of approximately $250\%$ relative to the as-received/control film. The S-passivated film shows an intermediate grain morphology, with a mean equivalent diameter of $31.6 \pm 9.0$~nm across 113 grains, corresponding to a smaller grain-size increase of approximately $82\%$. We use the AFM phase channel for this analysis because it provides enhanced contrast at grain boundaries relative to the topographic channel~\cite{pang2000phase}. Grain size and grain count are extracted from the phase-contrast images using Python~3.11 with the \texttt{scikit-image} image-processing library (\texttt{skimage}, version~0.22.0), and the resulting values are summarized in Table~\ref{tab:afm_summary}. More details regarding the extraction of grain size and grain density are provided in the Supplemental Materials.\

\begin{table}[htbp!]
    \centering
    \caption{Summary of AFM grain segmentation and surface roughness metrics for Nb films on differently prepared InP substrates.}
    \label{tab:afm_summary}
    \small
    \setlength{\tabcolsep}{4pt}
    \begin{tabular}{lcccc}
    \hline
    Sample & Mean $D_\mathrm{eq}$ & Median & Count & RMS \\
           & (nm) & (nm) &  & (nm) \\
    \hline
    As-received & $17.3 \pm 8.2$ & 16.7 & 440 & 2.1 \\
    Milled & $61 \pm 18.1$ & 59 & 57 & 3.5 \\
    S-passivated & $31.6 \pm 9.0$ & 31.4 & 113 & 0.6 \\
    \hline
    \end{tabular}
\end{table}

\begin{figure}[htbp!]
    \centering
    \includegraphics[width=0.9\linewidth]{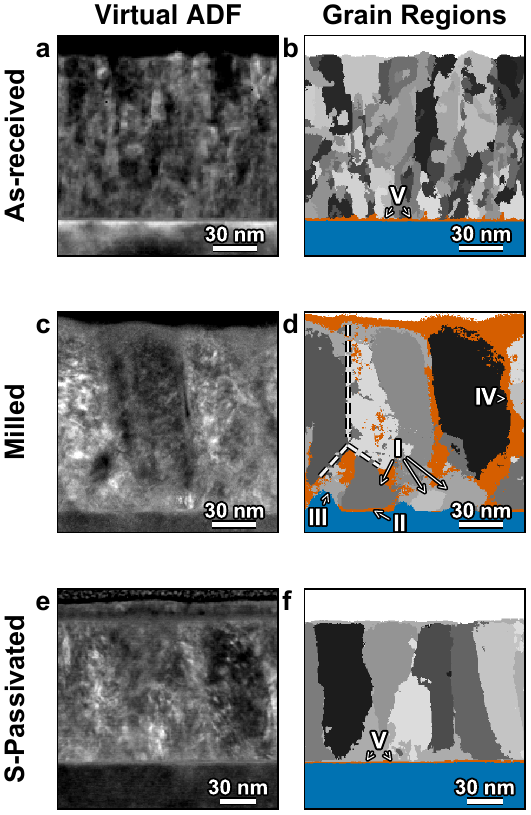}
    \caption{
    Virtual annular dark-field (ADF) images reconstructed from the 4D-STEM datasets and cluster-based grain maps representing the projected Nb grain structure of the \textbf{(a)}, \textbf{(b)} as-received, \textbf{(c)}, \textbf{(d)} milled, and \textbf{(e)}, \textbf{(f)} S-Passivated Nb/InP samples. Panels \textbf{(a)}, \textbf{(c)}, and \textbf{(e)} show the virtual ADF images reconstructed by integrating the corresponding 4D-STEM diffraction patterns over an annular detector range, while panels \textbf{(b)}, \textbf{(d)}, and \textbf{(f)} show the segmented structural maps. Grayscale regions correspond to different Nb grain regions, as defined from the associated diffraction patterns shown in the Supplementary Materials. See Figures~\ref{fig:s_as-received_4DSTEM}-\ref{fig:s_sulfur_4DSTEM} cluster-mean diffraction patterns and Appendix~\ref{sec:4D-STEM Grain Analysis} for their interpretation. The InP substrate is shown in blue, and amorphous regions are shown in orange. In the milled sample, the amorphous region forms channel-like pathways (IV) and partial interfacial separation regions (II), while in the as-received and S-passivated samples, the amorphous region appears as a thin, continuous buried interfacial layer between the Nb film and the substrate (V). 
    }
    \label{fig:Grains}
\end{figure}

The corresponding AFM topography data show that these changes in grain morphology are accompanied by substantial differences in surface roughness. The Ar-milled film has the largest RMS roughness, roughly $67\%$ higher than the as-received/control film, consistent with substantial grain coarsening and surface texturing induced by the argon milling process. This interpretation is consistent with prior work showing that Ar$^+$ sputtering of InP can drive nanoscale surface roughness evolution and pattern formation, as well as Nb thin-film studies showing that energetic growth conditions strongly influence grain size, grain shape, and surface roughness~\cite{frost2000inp,kittiwatanakul2018nb,gao2022nb}. In contrast, the S-passivated film has the lowest RMS roughness, reduced by approximately $71\%$ relative to the as-received/control film, indicating a finer and substantially smoother surface morphology than the Ar-milled film. Prior studies of ammonium sulfide treatment on InP show that S-passivation removes or suppresses native oxide formation, supporting the interpretation that the chemically passivated surface provides a more controlled interface for subsequent Nb growth~\cite{tian2014inp,lee2019facile}. Together, these AFM results suggest that argon milling promotes substantial grain growth and roughening of the deposited Nb film, while S-passivation limits this coarsening and preserves a smoother surface morphology.

Cross-sectional HAADF-STEM imaging and STEM-EDS mapping for the three substrate preparations are presented in Figure~\ref{fig:tem_8panel}. Focused Ion Beam (FIB) liftouts were prepared using the method described in Appendix \ref{TEM_Liftout}. The 3$\times$4 panel is organized by sample preparation, with the as-received, Ar$^+$-milled, and S-passivated samples shown in Figure~\ref{fig:tem_8panel}(a--d), (e--h), and (i--l), respectively. For each sample, the columns show the HAADF-STEM image, Nb EDS map, O EDS map, and substrate EDS map. STEM-EDS and 4D-STEM measurements were performed on a Thermo Fisher Scientific Spectra Ultra operated at 300~kV. STEM-EDS mapping was acquired at a camera length of 87~mm using a semi-convergence angle of 30~mrad, spot size 3, and a scan step size of 1.8~nm. STEM-EDS data were acquired and processed using Thermo Fisher Scientific Velox software. For EDS map processing, the selected line families were C~K, O~K, P~K, In~L, and Nb~L for the as-received and Ar$^+$-milled samples, and C~K, O~K, P~K, S~K, Ti~K, In~L, and Nb~L for the S-passivated sample.

Image-based analysis of the EDS maps was performed as described in the Supplementary Material and shown in Figure~\ref{fig:s_TEM_analysis}. Briefly, the Nb film was first extracted from the Nb elemental map and used to define all analysis regions. The near-surface region was defined as the top 25~nm of the Nb film, while the buried-interface region was defined as the bottom 15~nm adjacent to the Nb--InP interface. Oxygen-rich regions were identified statistically from the O intensity relative to the background in the InP substrate. High-disorder regions were defined from spatially coincident reductions in Nb EDS and STEM intensity. Because ADF/HAADF-STEM contrast depends on projected mass-thickness and atomic-number contrast~\cite{nellist2000ADF, macarthur2016ADF}, while STEM-EDS provides a local Nb elemental-intensity map with sensitivity to thickness and channeling effects~\cite{spurgeon2017STEMEDS}, this combined mask identifies regions with reduced Nb continuity. These regions are referred to here as high-disorder regions because of their reduced local Nb density/continuity and extended structural contrast~\cite{phillips2012LAADF, oveisi2019ADF}; this label avoids directly assigning them to grain boundaries, while noting that such regions are consistent with defect-rich intergranular regions in polycrystalline films~\cite{quirk2024GB, lee2026OGB}. The same masks, thresholds, and region definitions were applied to all samples to enable direct comparison.

For the as-received reference sample, Figure~\ref{fig:tem_8panel}(a--d) establishes the baseline Nb/InP interface morphology and oxygen distribution. The Nb film appears laterally continuous in the HAADF-STEM image. A strong oxygen signal is observed along the bottom of the Nb film in Figure~\ref{fig:tem_8panel}, indicating the presence of an interfacial oxide layer. Image analysis of the cross-sectional STEM/EDS data, detailed in Figure~\ref{fig:s_TEM_analysis} and summarized in Tables~\ref{tab:Results_area_fraction_metrics} and~\ref{tab:Results_regional_mean_oxygen_intensity_normalized}, shows that 7.8\% of the total Nb area in the as-received sample is classified as O-rich, with no high-disorder signatures detected in the analyzed bulk Nb region. The normalized O intensity values in Table~\ref{tab:Results_regional_mean_oxygen_intensity_normalized} for both the surface and buried-interface regions are defined as $1.0$ because this sample serves as the reference for the regional mean oxygen-intensity comparison. These baseline values provide the normalization reference against which the Ar$^+$-milled and S-passivated samples are compared.

For the Ar$^+$-milled sample, the regional mean O intensity values in Table~\ref{tab:Results_regional_mean_oxygen_intensity_normalized} are $0.9$ for the surface region and $0.8$ for the buried Nb--InP interface region, where both values are normalized to the corresponding regions in the as-received reference sample. Thus, relative to the as-received reference, the milled sample shows only a modest reduction in the regional mean O signal at both the surface and buried interface. However, the area-fraction metrics in Table~\ref{tab:Results_area_fraction_metrics} indicate substantial degradation in film quality: $24.9\%$ of the total Nb area is classified as O-rich, $9.8\%$ of the bulk Nb area is classified as high disorder, and $91.7\%$ of the high-disorder area is also classified as O-rich. This strong spatial overlap indicates that the high-disorder regions in the milled film are closely associated with O accumulation. The In/P EDS results representing the milled InP substrate surface in Figure~\ref{fig:tem_8panel}(h) reveal pronounced pyramidal facets consistent with \{111\} planes when viewed down the [110] direction on the InP~(001) surface at the buried interface. Such faceting and roughness development are consistent with prior reports showing that Ar$^+$ ion milling and sputtering can induce near-surface damage, roughening, and pattern formation in InP~\cite{pearton1990ion, frost2000inp}. These pyramidal features are spatially correlated with the observed microstructure: high-disorder features, indicated by dashed vertical lines, align with the valleys between adjacent pyramids and propagate through the full thickness of the Nb film.

\begin{table}[htbp!]
    \centering
    \caption{
    Regional mean oxygen intensity values from STEM-EDS image analysis, normalized to the corresponding as-received mean intensities. Values are dimensionless; the as-received sample is therefore 1.0 for each region. These values are relative image-contrast metrics and are not quantitative cross-sample oxygen concentrations.
    }
    \label{tab:Results_regional_mean_oxygen_intensity_normalized}
    \begin{tabular}{lccc}
    \hline
    Region & As-received & Milled & S-passivated \\
    \hline
    Surface region & 1.0 & 0.9 & 0.9 \\
    Buried interface region & 1.0 & 0.8 & 0.2 \\
    \hline
    \end{tabular}
\end{table}

\begin{table*}[htbp!]
    \centering
    \caption{Area fractions extracted from image segmentation of cross-sectional TEM images of the Nb films. Values indicate the percentage of the specified analysis region classified as satisfying each condition.}
    \label{tab:Results_area_fraction_metrics}
    \begin{tabular}{lccc}
    \hline
     & As-received & Milled & Sulfur-passivated \\
    \hline
    Fraction of total Nb area classified as oxygen-rich & 7.8\% & 24.9\% & 6.5\% \\
    Fraction of bulk Nb area classified as high disorder & 0.0\% & 9.8\% & 0.0\% \\
    Fraction of high disorder area classified as oxygen-rich & --- & 91.7\% & --- \\
    \hline
    \end{tabular}
\end{table*}

The HAADF-STEM image in Figure~\ref{fig:tem_8panel}(i) shows substantially improved Nb film morphology following S-passivation relative to Ar$^{+}$ milling, consistent with prior reports that S-based treatments suppress native oxide formation and improve the chemical and electrical quality of InP surfaces and interfaces~\cite{Tao1992,tian2014inp,lee2019facile,alian2011asv,xu2013inpinterface}. The Nb EDS map in Figure~\ref{fig:tem_8panel}(j) reveals a continuous film without visible low-signal gaps or pronounced high-disorder features, with no pixels in the analyzed interior Nb region exceeding the selected high-disorder threshold. The O EDS map in Figure~\ref{fig:tem_8panel}(k) also shows strongly suppressed O incorporation relative to the milled sample. Only $6.5\%$ of the total Nb area is classified as O-rich, compared with $24.9\%$ for the milled sample, corresponding to an approximately fourfold reduction in the projected area classified as O-rich under the applied criterion. The regional mean O intensities in Table~\ref{tab:Results_regional_mean_oxygen_intensity_normalized} further show normalized surface- and buried-interface-region values of $0.9$ and $0.2$, respectively, indicating that the buried-interface O signal is strongly reduced relative to the as-received reference. As shown in Figure~\ref{fig:tem_8panel}(l), the Nb--InP interface remains sharp and planar, with no evidence of the pyramidal faceting or etch-induced roughness produced by Ar$^{+}$ milling. Together, the HAADF-STEM and EDS results indicate that S-passivation preserves interface planarity while suppressing the structural defect pathways associated with O incorporation in the milled film~\cite{lee2026OGB}. The reduced O incorporation is also consistent with the ToF-SIMS depth profile in Figure~\ref{fig:sims_depth_profiles}(c).

Further structural insight is provided by the grain-resolved 4D-STEM analysis in Figure~\ref{fig:Grains}. For the as-received film, the virtual ADF image reconstructed from the 4D-STEM dataset in Figure~\ref{fig:Grains}(a) and the corresponding grain map in Figure~\ref{fig:Grains}(b), generated by clustering the diffraction patterns using the procedure described in Appendix~\ref{sec:4D-STEM Grain Analysis} and interpreted using the cluster-mean diffraction patterns shown in Figure~\ref{fig:s_as-received_4DSTEM}, reveal a highly fragmented polycrystalline microstructure. The film contains many small, laterally heterogeneous grain regions rather than a few extended columnar grains, indicating substantial orientational variation throughout its thickness. This fine-grained structure is consistent with the surface morphology observed by AFM in Figure~\ref{fig:afm_phase_topography}. A second defining feature is the laterally extended amorphous layer at the buried interface, highlighted by region V in Figure~\ref{fig:Grains}(b). This amorphous band separates the InP substrate from the overlying polycrystalline Nb, showing that the as-received film nucleates above a structurally disordered interface. The as-received sample therefore combines a fine-grained polycrystalline film with a distinct amorphous interfacial layer, whereas the treated samples in Figure~\ref{fig:Grains}(d,f) exhibit larger and more vertically continuous crystalline domains.

In the milled sample, the pyramidal InP surface structures observed in Figure~\ref{fig:tem_8panel}(h) are reproduced in the 4D-STEM segmentation in Figure~\ref{fig:Grains}(d)(III). Between adjacent pyramids, Figure~\ref{fig:Grains}(d)(I) identifies small Nb grain regions, some with projected areas of approximately $30~\mathrm{nm}^{2}$, separated from the larger grain regions by orange amorphous regions. The amorphous material only partially separates the Nb film from the substrate. In some locations, such as Figure~\ref{fig:Grains}(d)(II), it forms a continuous interfacial layer. At the pyramidal peaks in Figure~\ref{fig:Grains}(d)(III), however, crystalline Nb appears to approach the InP substrate more directly, with no clearly resolved amorphous interfacial layer at the spatial resolution of the 4D-STEM data. Within the peak-to-peak gaps in Figure~\ref{fig:Grains}(d)(IV), narrow vertical channels of amorphous material separate larger crystalline domains and form disordered pathways through the Nb film thickness. These channels follow the same spatial pattern as the locally reduced-Nb regions in Figure~\ref{fig:tem_8panel}(f) and the O-rich regions in Figure~\ref{fig:tem_8panel}(g), linking the extended structural disorder observed by 4D-STEM to the preferential O localization observed by EDS.

The S-passivated segmentation in Figure~\ref{fig:Grains}(f) exhibits a markedly different structure. Rather than forming intermittent interfacial regions and vertical channels as in the milled sample, the amorphous material is confined to a thin, laterally continuous layer that uniformly separates the crystalline Nb film from the InP substrate. Above this layer, the Nb film contains larger, continuous grain regions without the vertical amorphous pathways or high-disorder discontinuities observed after Ar$^{+}$ milling. This structure is consistent with the planar interface and absence of preferential bulk O segregation in the S-passivated Nb and O EDS maps in Figure~\ref{fig:tem_8panel}(j--k). The combined 4D-STEM and EDS results therefore show that S-passivation confines the amorphous component to the buried interface and prevents the vertically propagating, O-rich disorder produced by Ar$^{+}$ milling.

\begin{figure*}[htbp!]
    \centering
    \includegraphics[width=1\linewidth]{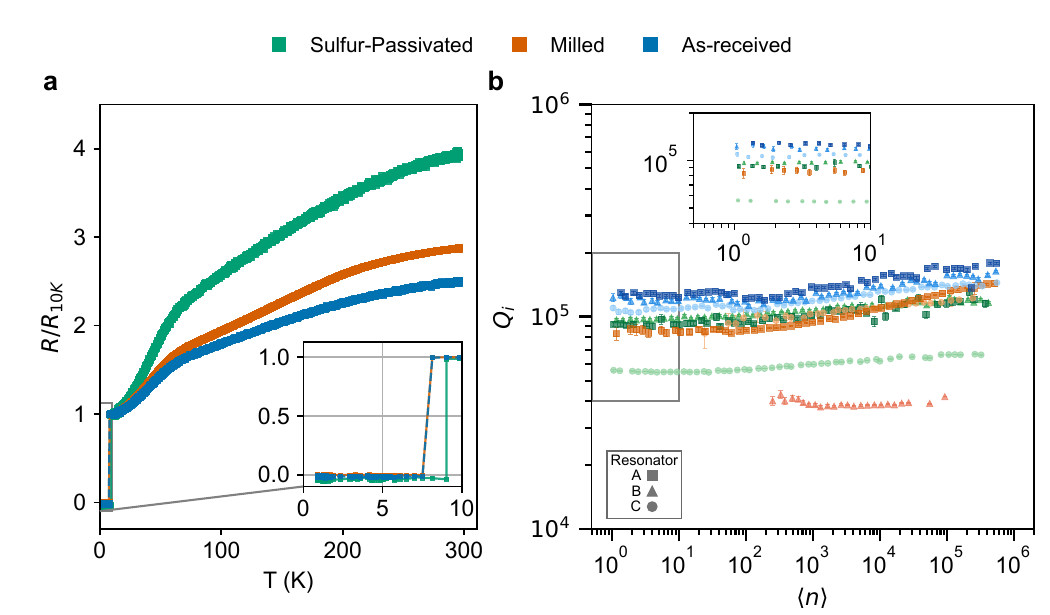}
    \caption{
    Superconducting $T_\textup{c}$ and internal microwave quality factor,
    $Q_\textup{i}$, for Nb films on InP~(001) with different substrate surface
    treatments.
    \textbf{(a)} Temperature-dependent normalized resistance,
    $R/R_{\textup{10K}}$, for S-passivated, milled, and as-received samples.
    \textbf{(b)} Internal quality factor, $Q_\textup{i}$, plotted as a function of
    average resonator photon occupation, $\bar{n}$, extracted from calibrated
    on-chip microwave power. Marker shapes labeled A--C identify the three
    resonators within each treatment. Resonators were fit using the complex
    diameter-corrected notch-resonator method described in
    Appendix~\ref{symetric_fitting}, except for the asymmetric milled
    resonator B, which was fit using the pole-zero
    method described in Sec.~\ref{asymetric_fitting}.
    }
    \label{fig:transport_qi}
\end{figure*}

Figure~\ref{fig:transport_qi}(a) shows the temperature-dependent normalized resistance of the Nb films, where each trace is normalized to its normal-state resistance measured between 10 and 12~K. The S-passivated film exhibits the highest superconducting transition temperature, with $T_c \approx 9.0~\mathrm{K}$, approximately 1~K higher than the milled and as-received films, which, within the measurement resolution $\approx 0.2~\mathrm{K}$, share a common transition near $8.1~\mathrm{K}$.

The S-passivated Nb film shows the highest room-temperature normalized resistance ratio, reaching $R/R_N\approx4$ at 296~K, compared with $\approx2.9$ for the milled film and $\approx2.5$ for the as-received film. Relative to the as-received case, argon milling provides a modest improvement, while S-passivation results in a much higher $R/R_N$. A higher room-temperature resistance ratio is consistent with reduced impurity and defect scattering in the film, indicating improved crystalline quality and fewer scattering sites such as grain boundaries and incorporated O.

CPW Resonators were fabricated using the methods described in Appendix \ref{Res_fab}. Microwave measurements were then preformed in a Bluefors SD dilution fridge at 20 mK as described in Appendix \ref{Resonator_measurement}. Two fitting procedures were used, described in Appendix \ref{fitting}, to extract the resonator parameters shown in
Fig.~\ref{fig:transport_qi}(b). The photon number was calculated from the fitted resonator parameters and the estimated microwave power delivered on chip.

Figure~\ref{fig:transport_qi}(b) shows the resulting internal quality factor, $Q_i$, as a function of photon number for the three surface preparations. At the lowest measured powers, the as-received resonators give the largest $Q_i$ values, with $Q_i \approx 1.1$--$1.3\times10^{5}$ across the three resonator frequencies. The S-passivated resonators are somewhat lower, with two resonators near $9\times10^{4}$--$1.0\times10^{5}$ and one lower-$Q_i$ resonator near $5.6\times10^{4}$. The milled sample shows a similar spread, with two resonators near $8.3\times10^{4}$ and the asymmetric $5.04~\mathrm{GHz}$ resonator reduced to approximately $4.0\times10^{4}$.

These data therefore do not show a simple improvement of microwave loss with the surface treatments that improved the structural or chemical metrics discussed above. Instead, the as-received sample performs at least as well as, and generally better than, the S-passivated and milled samples over the measured photon-number range. This suggests that the metal--substrate interface modified by these preparations is not the dominant loss channel in the present resonator/package system. Rather, the microwave loss may be limited by other contributions, such as substrate bulk loss, package or radiation loss, or quasiparticle loss. The relatively weak power dependence of $Q_i$ also suggests that the devices are not strongly limited by saturable two-level-system dielectric loss in the measured range. Thus, while aggressive surface preparation can clearly degrade individual resonators, especially for the milled sample, the present data do not support a conclusive reduction of microwave loss from S passivation or milling alone.

\section{Conclusion}
In conclusion, we presented a systematic study of how buried Nb--InP interface preparation affects the chemistry, structure, superconducting transport, and microwave loss of sputtered Nb resonators. Comparing an as-received control, \textit{in-situ} Ar$^{+}$ milling, and sulfur passivation, we find that removing or modifying the native interfacial oxide can produce a cleaner and sharper metal--semiconductor interface and, in the sulfur-passivated case, improved superconducting transport. However, these improvements do not translate into improved microwave performance. Because sulfur is detected in all samples, the degraded performance of the sulfur-passivated devices cannot be attributed simply to the introduction of sulfur as a more lossy interfacial TLS than oxygen. Likewise, the available structural and morphological characterization does not support an explanation based on dramatic Nb damage or large-scale roughening. Instead, our results indicate that microwave loss in Nb--InP devices is not limited solely by amorphous interfacial oxide, and that removing this oxide may expose or enhance an additional loss channel associated with the metal--semiconductor interface or the InP substrate. One possible mechanism is electromechanical or piezoelectric loss in the III--V substrate, with the native oxide acting as a partial interfacial buffer, although this interpretation remains speculative and requires further study. These findings are important for hybrid superconductor--semiconductor quantum devices because they show that buried-interface engineering must be evaluated by microwave loss directly, not only by chemical sharpness, oxide removal, or superconducting transport metrics.

\subsection*{Acknowledgments}

We acknowledge support from the Defense Advanced Research Projects Agency (DARPA) Synthetic Quantum Nanostructures (SynQuaNon) program under Grant Agreement No. HR00112420343. We also acknowledge support from National Science Foundation Clemson (2137776). 
The STEM/EDX and ToF-SIMS measurements and analyses were supported by the U.S. Department of Energy, Office of Science, National Quantum Information Science Research Centers, Co-design Center for Quantum Advantage (C2QA) under Contract No. DE-SC0012704 (PNNL FWP 76274). A portion of this research was performed on a project award (10.46936/staf.proj.2026.62105/60015729) from the Environmental Molecular Sciences Laboratory, a DOE Office of Science User Facility sponsored by the Biological and Environmental Research program under Contract No. DE-AC05-76RL01830.

The authors would like to gratefully acknowledge Bethany Matthews for assistance with STEM sample preparation.

\subsection*{Data Availability}
All data supporting the findings of this study are available from the corresponding author upon request.

\beginsupplement
\appendix

\setcounter{figure}{0}
\setcounter{table}{0}

\section{ToF-SIMS Methods}

\subsection{Data Acquisition}
\label{SIMS_Measurement}

A TOF.SIMS 5 instrument (IONTOF GmbH, M\"unster, Germany) was used in dual-beam interlaced depth-profiling mode. A 2.0~keV Cs$^{+}$ beam was used for sputtering, and a 25~keV Bi$^{+}$ beam was used as the analysis beam for secondary-ion signal collection. The Cs$^{+}$ sputtering beam was rastered over a $300 \times 300~\mu\mathrm{m}^{2}$ area. The Bi$^{+}$ analysis beam was focused to an approximately $5~\mu\mathrm{m}$ diameter spot, with a beam current of approximately 1.4~pA and a repetition frequency of 20~kHz. The Bi$^{+}$ beam was rastered over a $100 \times 100~\mu\mathrm{m}^{2}$ area centered within the Cs$^{+}$ sputter crater. SurfaceLab software, version 7.2, provided by the ToF-SIMS instrument manufacturer, was used to extract the relevant ion images and depth profiles.

\subsection{Interface Determination and Analysis}
\label{SIMS_Interfaces}

\begin{figure*}[htbp!]
    \centering
    \includegraphics[width=0.9\linewidth]{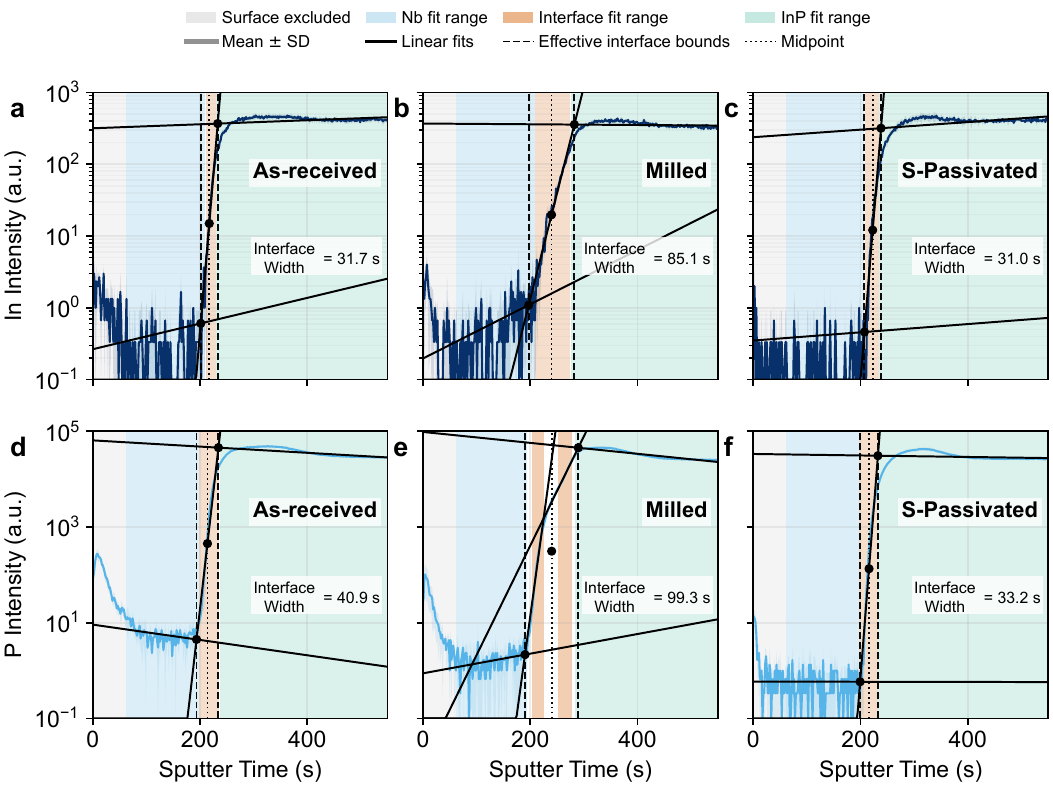}
    \caption{
    Piecewise-linear construction used to define the effective Nb--InP interface region from In and P ToF-SIMS depth profiles. Panels (a)--(c) show the In-based construction for the as-received, Ar-milled, and S-passivated samples, respectively, and panels (d)--(f) show the corresponding P-based construction. The extracted upper and lower bounds define the apparent interface width.
    }
    \label{fig:s_sims_interface_method}
\end{figure*}

The effective Nb--InP interface was analyzed using two parallel, indicator-specific applications of the same fitting procedure, one based on the In depth profile and the other on the P depth profile, to test whether the conclusions depend on the choice of substrate-species indicator. As shown in Figure~\ref{fig:s_sims_interface_method}, linear fits were performed in log-intensity space over sputter-time windows representing the Nb-side baseline, one or more interface-transition segments, and the substrate-side InP baseline; the first $60~\mathrm{s}$ of each profile were excluded from the Nb-side fit to avoid weighting the construction by the oxidized Nb--air surface. The independently fitted In-derived $t_{\mathrm{III}}$ and $t_{\mathrm{IV}}$ bounds were used consistently both to define the integration windows for the normalized $^{18}\mathrm{O}$:In, C:In, and S:In ratios and to calculate the apparent interface widths reported in Table~\ref{sims_tab_in}. The complete analysis was repeated using the independently fitted P-derived bounds, which were used for both the $^{18}\mathrm{O}$:P and C:P integrations and the P-derived interface widths reported in Figure~\ref{fig:s_sims_interface_method} and Table~\ref{s_P_sims_tab}. The In-based results are used for the main-text discussion because the In signal provides a more stable and sharply defined interface marker under the present sputtering conditions, whereas the P transition is more susceptible to broadening from preferential sputtering and/or the loss of volatile P-containing species. The P-based analysis therefore serves as a robustness check: the two indicators reproduce the same treatment-dependent trends, although their normalized ratios and apparent interface widths are not expected to be numerically identical.

For profiles with a single transition slope, the film-side interface bound, $t_{\mathrm{III}}$, was defined by the intersection of the Nb-side baseline fit with the transition fit, and the substrate-side bound, $t_{\mathrm{IV}}$, was defined by the intersection of the transition fit with the substrate-side baseline fit. The apparent interface width was calculated as
\[
\Delta t_{\mathrm{int}} = t_{\mathrm{IV}}-t_{\mathrm{III}}.
\]
The Ar-milled P profile required two transition fits: a gradual fit for the film-side bound and a steeper fit for the substrate-side bound. Approximate widths in nanometers were obtained by assigning the midpoint of the III--IV interval to the bottom of the $130~\mathrm{nm}$ Nb film.

The P-derived bounds were then used to calculate the P-normalized impurity metrics in Table~\ref{s_P_sims_tab}. For $X = {}^{18}\textup{O}$ or C, the numerator was integrated from the beginning of the sputter profile through the P-derived substrate-side bound, and the P denominator was integrated over the following $200~\mathrm{s}$ of the InP substrate:
\[
R_{X:\mathrm{P}} =
\frac{
\displaystyle\int_{t_0}^{t_{\mathrm{IV}}^{(\mathrm{P})}}
I_X(t)\,dt
}{
\displaystyle\int_{t_{\mathrm{IV}}^{(\mathrm{P})}}^{
t_{\mathrm{IV}}^{(\mathrm{P})}+200~\mathrm{s}}
I_{\mathrm{P}}(t)\,dt
}.
\]
The resulting $^{18}\textup{O}$:P and C:P ratios were normalized to the corresponding as-received values.

The P-based results in Table~\ref{s_P_sims_tab} are included as a consistency check on the In-based analysis used for the main text. They reproduce the same trends: Ar milling gives the broadest apparent interface and S passivation gives the lowest normalized oxygen and carbon incorporation. The absolute P-derived widths are somewhat larger and more fit-window-sensitive than the In-derived values, likely because preferential sputtering or desorption of the more volatile P component broadens the P transition during ion bombardment~\cite{petit2009inp}. For this reason, the In signal was used as the primary substrate indicator in Figure~\ref{fig:sims_depth_profiles}(d) and Table~\ref{sims_tab_in}.

\begin{table*}[htbp!]
    \centering
    \small
    \setlength{\tabcolsep}{6pt}
    \caption{
    Normalized integrated impurity-to-substrate-indicator ratios and apparent Nb--InP interface widths obtained from the TOF-SIMS depth profiles. Here, $X$ denotes the In or P substrate indicator listed in each subcolumn. For each indicator, the corresponding interface bounds were used consistently to calculate both the impurity ratios and the interface widths. Each impurity-to-$X$ ratio is normalized to its corresponding as-received value.
    }
    \label{s_P_sims_tab}
    \begin{tabular}{lcccccc}
    \hline
    & \multicolumn{2}{c}{As-received}
    & \multicolumn{2}{c}{Ar-milled}
    & \multicolumn{2}{c}{S-passivated} \\
    \cline{2-3}\cline{4-5}\cline{6-7}
    Metric & In & P & In & P & In & P \\
    \hline
    Normalized $^{18}\mathrm{O}:X$ & 1.00 & 1.00 & 0.54 & 0.55 & 0.03 & 0.03 \\
    Normalized $\mathrm{C}:X$     & 1.00 & 1.00 & 0.62 & 0.62 & 0.03 & 0.03 \\
    Normalized $\mathrm{S}:X$     & 1.00 & 1.00 & 0.51 & 0.51 & 0.52 & 0.41 \\
    Apparent interface width (s)   & 32   & 41   & 85   & 99   & 31   & 33   \\
    Approx. interface width (nm)   & 19   & 25   & 46   & 54   & 18   & 20   \\
    \hline
    \end{tabular}
\end{table*}

\section{AFM Phase Grain Size Analysis}

\begin{figure*}[htbp!]
    \centering
    \includegraphics[width=0.9\linewidth]{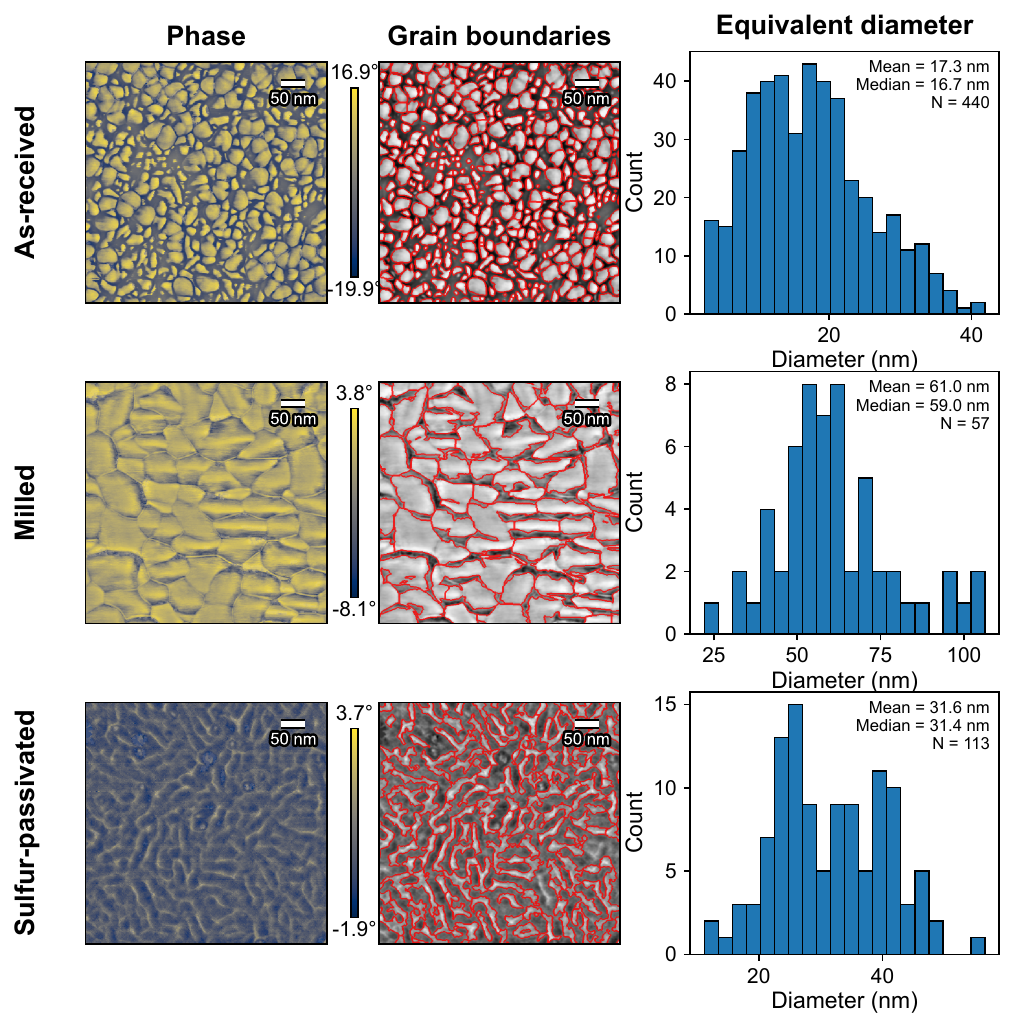}
    \caption{Summary of AFM phase-based grain segmentation analysis for the as-received, milled, and sulfur-passivated niobium films on InP~(001). Left column: processed phase images following background flattening, and, for the sulfur passivated case, an additional dust masking filter. Center column: extracted grain boundaries overlaid on the phase contrast using watershed-based segmentation. Right column: histograms of equivalent circular grain diameters extracted from the segmented grain areas. }

    \label{fig:s_grains}
\end{figure*}

\begin{figure*}[htbp!]
    \centering
    \includegraphics[width=0.9\linewidth]{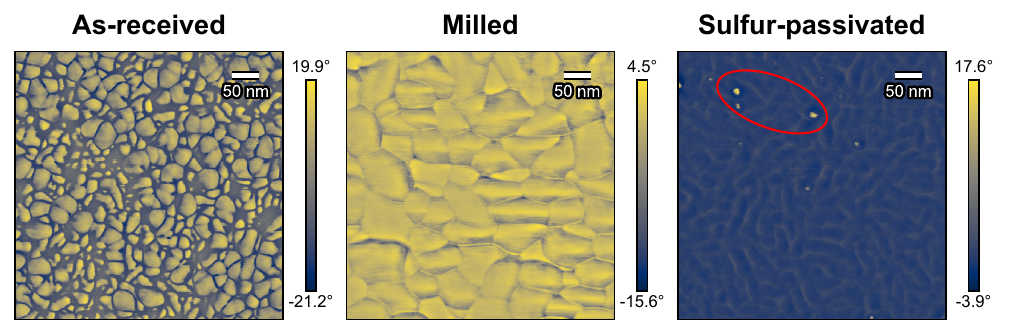}
    \caption{
    Phase images prior to dust-masking. A region in the sulfur passivated phase plot is encircled to show the dust particles that were masked in the grain analysis.  
    }
    \label{fig:s_grains_premask}
\end{figure*}

To provide a quantitative comparison of surface morphology across substrate preparation conditions, grain size analysis was performed on the AFM phase images shown in Figure~\ref{fig:s_grains}. Phase images were selected for this analysis because they provided stronger grain boundary contrast than the corresponding topography channels.

Each phase image was first background-flattened by subtracting a Gaussian-blurred version of the image in order to remove long-wavelength scan artifacts and increase local grain-scale contrast. For the sulfur-passivated sample, an additional outlier suppression mask was applied to remove isolated high-contrast dust features that were visually identified as non-physical surface artifacts. These dust features are identified in Figure~\ref{fig:s_grains_premask}. The dust correction resulted in a 1.4~nm or 4\% increase in grain size.

Grain regions were then segmented using threshold-based binary masking. "Bright" regions, those with higher phase magnitude, were classified as grains. A distance-transform watershed algorithm was then applied to separate neighboring grains sharing contiguous boundaries, implemented in Python 3.11 using the \texttt{watershed} routine from the \texttt{scikit-image} (\texttt{skimage}, version 0.22.0) library. Grain boundaries were extracted from the labeled segmentation map and overlaid on the processed phase images for visual verification.

For each segmented grain, the equivalent circular diameter was calculated from the enclosed pixel area according to
\[
d_{\mathrm{eq}} = 2\sqrt{\frac{A}{\pi}},
\]
where \(A\) is the grain area. The resulting histograms summarize the distribution of equivalent grain diameters for each surface preparation condition

\begin{figure*}[htbp!]
    \centering
    \includegraphics[width=7in]{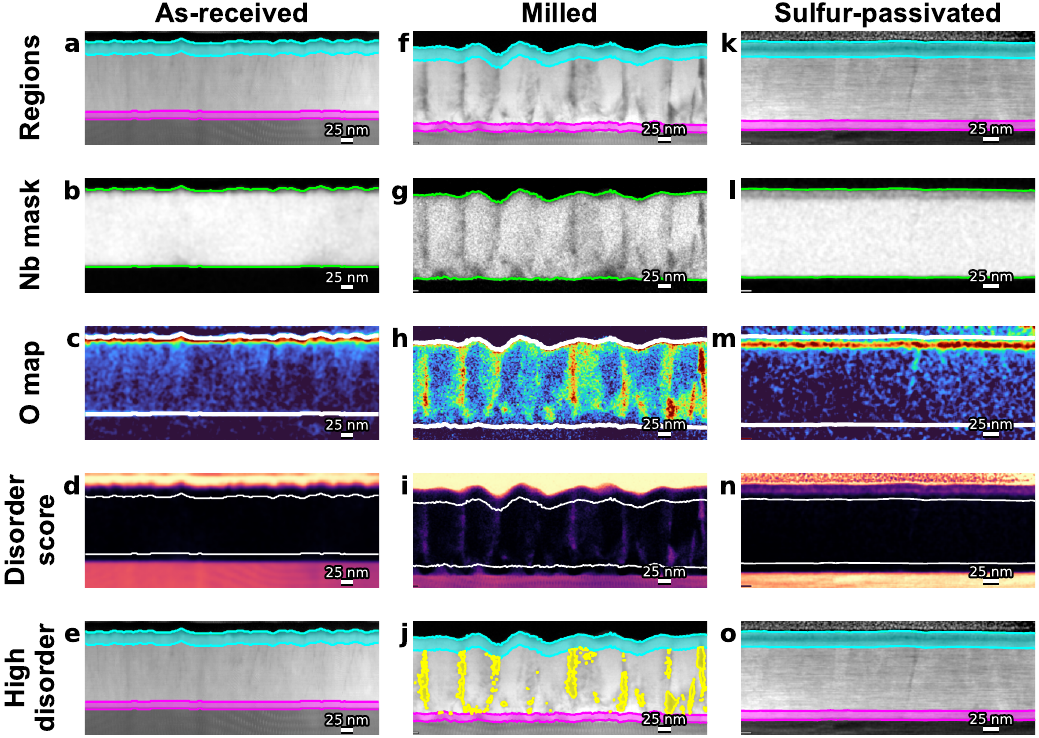}
    \caption{
    Image-based regional analysis of cross-sectional scanning transmission electron microscopy (STEM) and energy-dispersive X-ray spectroscopy (EDS) data for niobium films deposited on differently prepared InP substrates. The left column (\textbf{a--e}) corresponds to the as-received sample, the middle column (\textbf{f--j}) corresponds to the argon-milled sample, and the right column (\textbf{k--o}) corresponds to the sulfur-passivated sample. Each row compares the same analysis step across the three samples, with row labels shown at left. \textbf{(a,f,k)} Segmented near-surface and buried-interface regions of the Nb film used for regional oxygen analysis; the top surface region is highlighted in cyan and the buried Nb--InP interface region in magenta. \textbf{(b,g,l)} Processed Nb signal with the extracted Nb film mask overlaid as green lines, defining the full analysis region used for thickness and area-based measurements. \textbf{(c,h,m)} Background-corrected oxygen intensity map shown in false color, with the Nb mask outline overlaid as white lines. \textbf{(d,i,n)} Nb/STEM agreement score map used to identify candidate high-disorder and low-density regions, where reduced Nb signal and reduced STEM intensity spatially coincide. White lines indicate the bottom of the surface region and the top of the buried-interface region. \textbf{(e,j,o)} Final high-disorder mask overlaid on the STEM image, highlighting regions of likely Nb discontinuity and/or reduced crystallinity in yellow. These high-disorder regions are used for analysis of preferential oxygen segregation. The surface region is shown in cyan and the buried-interface region in magenta.
    }
    \label{fig:s_TEM_analysis}
\end{figure*}

\section{TEM/EDS}

\subsection{FIB Liftout Method}
\label{TEM_Liftout}

Cross-sectional TEM specimens were prepared from three $130~\mathrm{nm}$ Nb films grown on InP~(001) substrates subjected to three different surface treatments. All three specimens were prepared by focused ion beam (FIB) lift-out, with the lift-out regions selected away from craters generated during previous time-of-flight secondary ion mass spectrometry (ToF-SIMS) measurements. The cross-sectional lamellae were prepared using a Thermo Fisher Scientific Helios Hydra UX DualBeam FIB--SEM. Prior to ion milling, the sample surface was protected by an electron-beam-deposited carbon coating approximately $100$--$500~\mathrm{nm}$ thick, followed by an approximately $1~\mu\mathrm{m}$-thick plasma-beam-deposited Pt-containing protective coating. The lamellae were then lifted out and progressively thinned to electron transparency using reduced ion-beam currents. Final polishing was performed at $5~\mathrm{kV}$ and $30~\mathrm{pA}$ to reduce FIB-induced surface damage and amorphization.

\subsection{EDS Processing and Regional Thresholding Methodology}

All image-based regional analysis shown in Figure~\ref{fig:s_TEM_analysis} was performed in Python using \texttt{numpy}, \texttt{matplotlib}, \texttt{scipy}, and \texttt{scikit-image}. Functions from \texttt{scikit-image} were used for Gaussian filtering, binary filtering, connected-component analysis, and mask generation. These routines were used to define spatial regions within the Nb film for quantitative oxygen-intensity analysis.

The spatial calibration of each cross-sectional dataset was determined from the TEM scale bar. For each sample, the pixel length corresponding to 50~nm was measured directly from the image, and all region thicknesses and area calculations were converted from pixels to physical dimensions using the corresponding nm/pixel calibration factor. This calibration was performed independently for the as-received, argon-milled, and sulfur-passivated datasets to account for differences in image magnification.

The full Nb film region was extracted from the processed Nb elemental intensity map. The Nb intensity image was Gaussian smoothed and rescaled using percentile normalization. The film mask was then defined by thresholding the normalized Nb intensity using a fixed relative threshold. Connected-component filtering, small-object removal, and hole filling were applied to retain the continuous Nb film region. The upper and lower Nb film boundaries were extracted as a function of lateral position from this mask, as shown in Figure~\ref{fig:s_TEM_analysis}(b), (g), and (l).

The near-surface and buried-interface regions were defined directly from the extracted Nb film boundaries as a function of lateral position. The surface region was defined as the top 25~nm of the Nb film:
\[
t_{\mathrm{surface}} = 25~\mathrm{nm},
\]
while the buried-interface region was defined as the bottom 15~nm of the film adjacent to the Nb--InP interface:
\[
t_{\mathrm{interface}} = 15~\mathrm{nm}.
\]
The remaining interior film volume was classified as the bulk Nb region. The near-surface, bulk, and buried-interface regions obtained from these boundaries are shown in Figure~\ref{fig:s_TEM_analysis}(a), (f), and (k).

Oxygen intensities were corrected for image-specific background using the InP substrate region below the Nb--InP interface. The oxygen map was first Gaussian smoothed and percentile normalized. For each lateral position, the lower Nb film boundary extracted from the Nb mask was used to define the substrate background region as all pixels below the buried-interface boundary. The mean substrate oxygen signal, $\mu_{\mathrm{sub}}$, was subtracted from the processed oxygen map:
\[
I_{\mathrm{O,corr}} = \max\left(I_{\mathrm{O}}-\mu_{\mathrm{sub}},0\right).
\]
Negative values after subtraction were clipped to zero.

Oxygen-rich pixels were defined statistically from the background-corrected oxygen map. Pixels were classified as oxygen-rich when
\[
I_{\mathrm{O,corr}} > 3\sigma_{\mathrm{sub}},
\]
where $\sigma_{\mathrm{sub}}$ is the standard deviation of the oxygen intensity measured in the InP substrate background region. Regional mean oxygen intensities were computed from $I_{\mathrm{O,corr}}$ within the surface, bulk, buried-interface, and high-disorder regions. For tabulated comparison, these background-corrected mean intensities were normalized to the corresponding as-received region mean. Because the EDS maps were independently rescaled, the regional mean O values are reported as relative image-contrast metrics and are not interpreted as quantitative cross-sample oxygen concentrations.\\

To identify regions of likely reduced film continuity and grain-boundary-associated disorder, intermediate low-intensity masks were first constructed from the processed Nb and STEM images. Candidate low-Nb pixels were identified within the Nb film, excluding the near-surface region, as pixels satisfying
\[
I_{\mathrm{Nb}} < 0.60\,I_{95,\mathrm{Nb}},
\]
where $I_{95,\mathrm{Nb}}$ is the 95th percentile Nb intensity measured within the full film mask. Similarly, candidate low-STEM pixels were identified within the Nb film, excluding the near-surface region, according to
\[
I_{\mathrm{STEM}} < 0.60\,I_{95,\mathrm{STEM}},
\]
where $I_{95,\mathrm{STEM}}$ is the 95th percentile STEM intensity inside the Nb film mask. These low-intensity masks were used only as intermediate criteria for identifying high-disorder regions and were not treated as separately reported material regions.\\

A composite Nb/STEM agreement score was then calculated from the processed Nb and STEM intensity maps:
\[
S_{\mathrm{comp}} = \left(1-I_{\mathrm{Nb}}\right)\left(1-I_{\mathrm{STEM}}\right).
\]
Thus, pixels with simultaneously reduced Nb signal and reduced STEM intensity yield high values of $S_{\mathrm{comp}}$. The score was evaluated within the bulk Nb region, excluding the near-surface and buried-interface bands, as shown in Figure~\ref{fig:s_TEM_analysis}(d), (i), and (n). The final high-disorder mask was generated by thresholding this composite score and requiring agreement with a loose low-intensity gate in either the Nb or STEM channel. Binary filtering was then applied to remove isolated pixels and fill small voids.

The resulting high-disorder masks are shown in Figure~\ref{fig:s_TEM_analysis}(e), (j), and (o). These regions are interpreted as areas of likely Nb discontinuity, reduced local Nb density, and/or reduced crystallinity, and were used for quantification of preferential oxygen segregation. Local variations in projected thickness and diffraction/channeling may also affect the Nb-EDS and HAADF-STEM intensities. The resulting O-rich and high-disorder area fractions are therefore interpreted as relative, threshold-dependent projected-area metrics rather than absolute phase or defect-volume fractions. A value of 0\% means that no pixels exceeded the selected threshold in the analyzed field of view.

Although the threshold values are empirically selected, all threshold definitions, percentile normalizations, background corrections, and region thicknesses were applied identically across the as-received, argon-milled, and sulfur-passivated datasets to ensure direct and internally consistent comparison.

\section{4D-STEM Acquisition and Cluster-based Grain Analysis}
\label{sec:4D-STEM Grain Analysis}

Four-dimensional scanning transmission electron microscopy (4D-STEM) datasets were acquired using an EMPAD detector at a camera length of $550~\mathrm{mm}$. The microscope was operated in microprobe mode with spot size 9, giving a probe semi-convergence angle of approximately $2~\mathrm{mrad}$. A scan step size of $1.5~\mathrm{nm}$ was used for all 4D-STEM acquisitions. This microprobe condition was selected to obtain small-convergence diffraction patterns while retaining nanometer-scale spatial sampling suitable for grain mapping. Virtual ADF images were reconstructed by integrating the diffraction intensity over an annular range of 20-40 mrads at each scan position.

The diffraction-pattern datasets were analyzed using principal component analysis (PCA) denoising, non-negative matrix factorization (NMF), and clustering. In this workflow, each cluster represents a class of scan positions with similar diffraction patterns. PCA denoising was performed using 30 retained components for all analyzed datasets.

Because the Milled and S-Passivated samples exhibited larger grains, whereas the as-received sample showed a substantially finer-grained structure and greater local diffraction-pattern complexity, different NMF and clustering parameters were used for the different samples. For the Milled and S-Passivated samples, 12 NMF components and 12 clusters were used. For the as-received sample, 30 NMF components and 15 clusters were used. Because these parameters differed among the datasets, differences in apparent grain size in the resulting 4D-STEM maps are interpreted qualitatively. Independent AFM analysis provides complementary quantitative measurements of the surface grain size.

After clustering, grain regions were operationally defined as spatially connected components within each cluster map; disconnected regions with the same cluster label were counted as separate grain regions. Representative mean diffraction patterns were then calculated to examine the diffraction features associated with the mapped regions. Because the milled and S-passivated samples contained fewer grain regions, their diffraction patterns were averaged by cluster; for the as-received sample, which contained more grain regions, they were averaged separately within each grain region. During post-processing, grain regions smaller than $25~\mathrm{nm}^{2}$ were merged with adjacent regions based on diffraction-pattern similarity.

Across all three samples, most averaged diffraction patterns are dominated by one set of reflections, supporting assignment of the corresponding connected map regions as grain regions. Additional reflections may arise from grain overlap along the electron-beam direction or from double diffraction. A few patterns lack a dominant set; their corresponding regions are described as having mixed diffraction features. The maps therefore represent the projected grain structure rather than individual grains throughout the lamella thickness.

\begin{figure*}[htbp!]
    \centering
    \includegraphics[width=\linewidth]{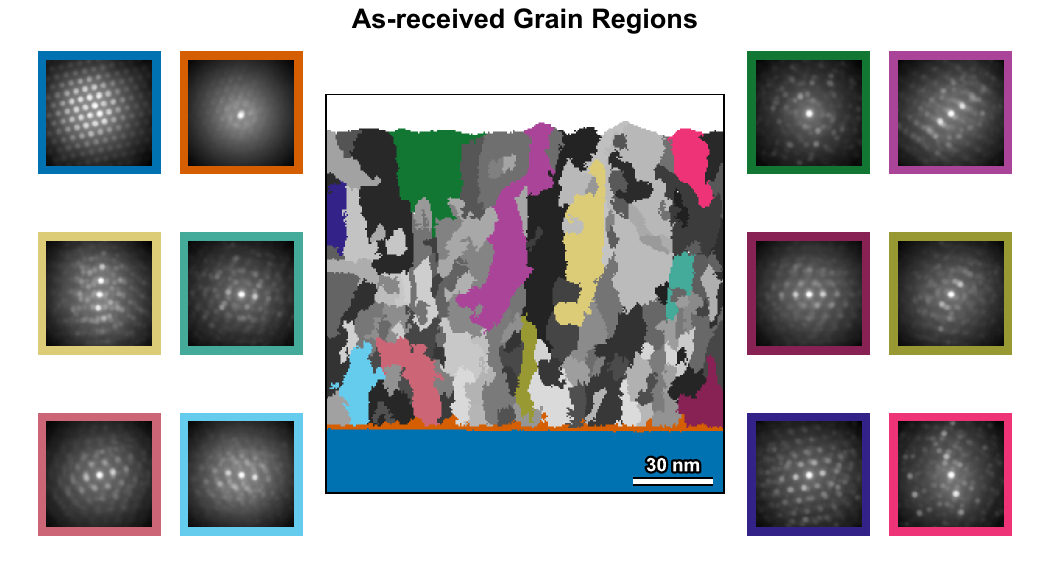}
    \caption{
    4D-STEM cluster-based grain map of the projected Nb grain structure and representative grain-region-averaged diffraction patterns for the as-received Nb/InP sample. The colored frames around each diffraction pattern match the grain-region colors in the map. The displayed diffraction patterns are averaged within each segmented grain region and shown on a logarithmic intensity scale. Log scaling enhances weak reflections and helps distinguish crystalline, amorphous, and mixed diffraction features. The as-received patterns are generally more complex than those of the treated samples. Most are dominated by one set of reflections, while a few show mixed diffraction features; weaker additional reflections may arise from projected grain overlap or double diffraction. The orange region corresponds to the amorphous component observed near the buried interface while the blue region corresponds to the crystalline InP~(001) substrate.
    }
    \label{fig:s_as-received_4DSTEM}
\end{figure*}

\begin{figure*}[htbp!]
    \centering
    \includegraphics[width=\linewidth]{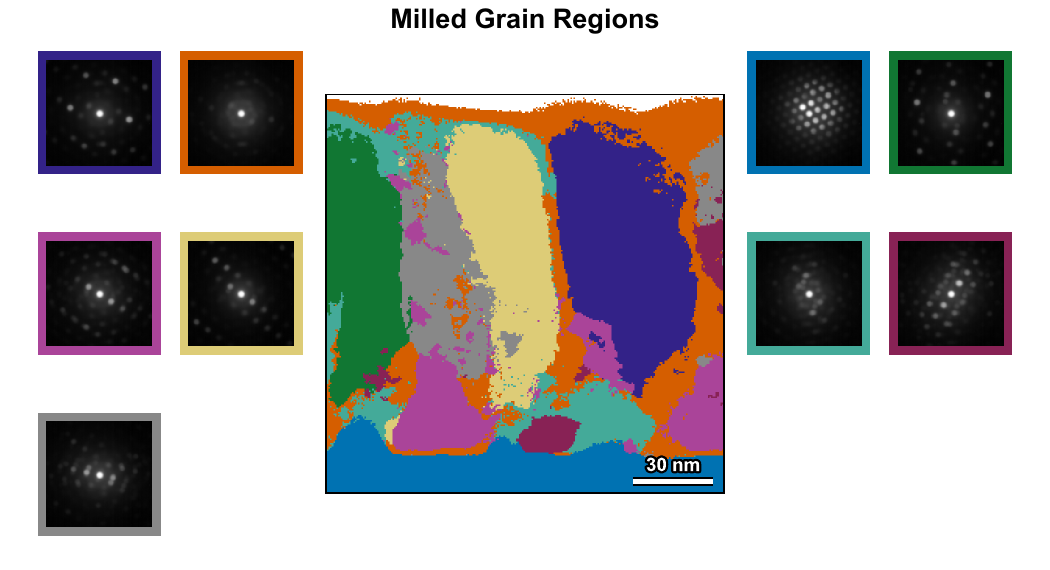}
    \caption{
    4D-STEM grain-region map and corresponding cluster-averaged diffraction patterns for the milled Nb/InP sample. See Figure~\ref{fig:s_as-received_4DSTEM} for the display scaling and Appendix~\ref{sec:4D-STEM Grain Analysis} for the averaging procedure and diffraction-pattern interpretation. The orange region corresponds to the amorphous component observed near the buried interface, Nb-cap interface, and connecting high disorder channels, while the blue region corresponds to the crystalline InP~(001) substrate.
    }
    \label{fig:s_milled_4DSTEM}
\end{figure*}

\begin{figure*}[htbp!]
    \centering
    \includegraphics[width=\linewidth]{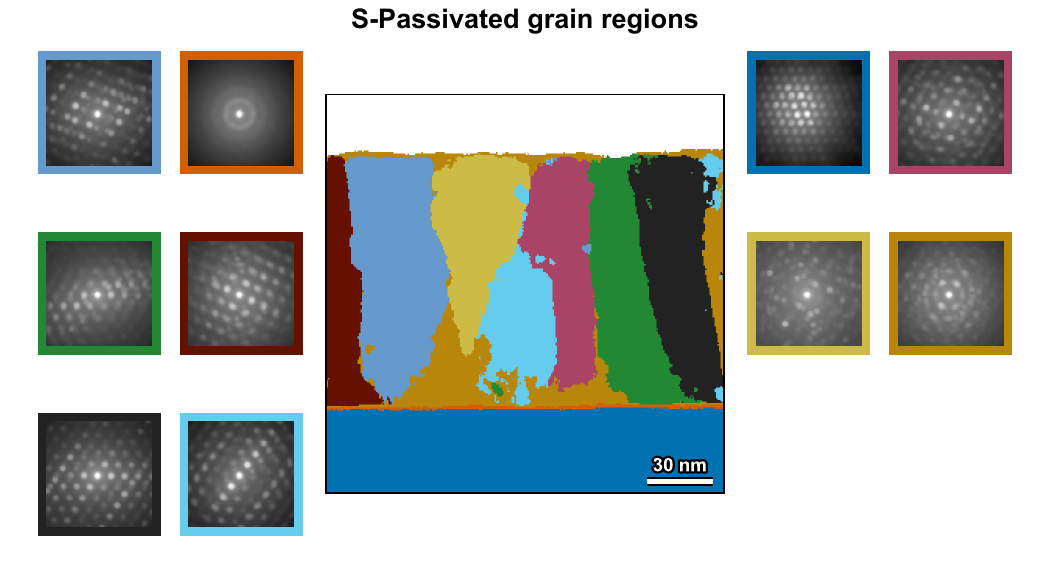}
    \caption{
    4D-STEM grain-region map and corresponding cluster-averaged diffraction patterns for the sulfur-passivated Nb/InP sample. See Figure~\ref{fig:s_as-received_4DSTEM} for the display scaling and Appendix~\ref{sec:4D-STEM Grain Analysis} for the averaging procedure and diffraction-pattern interpretation. 
    }
    \label{fig:s_sulfur_4DSTEM}
\end{figure*}

\subsection{Electromagnetic Simulation and Resonator Design}

Electromagnetic simulations were performed in \textit{Cadence AWR AXIEM} to design the inductively coupled coplanar waveguide (CPW) resonators used in this work and to determine the resonator geometries required to achieve the target resonance frequencies and external coupling quality factors. Simulations were carried out on individual resonators prior to assembly into the final resonator layout.

For each resonator, the generated GDS layout was imported into AXIEM as a planar electromagnetic structure. Launch pads were removed prior to simulation so that the EM solve contained only the straight CPW feedline section and the inductively coupled resonator geometry. The final design utilized a feedline CPW geometry consisting of a $42~\mu$m center trace with $25~\mu$m trenches, while the resonator CPW geometry used a $35~\mu$m center trace with $20~\mu$m trenches. The substrate was modeled as InP with relative dielectric constant $\varepsilon_r = 12.4$, and the Nb metallization layer was treated as a perfect conductor.

\subsubsection{Port Configuration}

Single-resonator EM simulations were implemented as a three-port differential-face structure \cite{smitham2025modeling}. Two ports were placed on the left and right feedline cross-sections, corresponding to the feedline input and output, respectively. A third differential-face port was placed at the far open end of the resonator to directly probe the resonator admittance response.

Each differential-face port was defined on the Nb layer with the positive terminal assigned to the CPW center conductor and the negative terminals assigned to the adjacent ground-plane trench edges. The resonator end nearest the transmission line was shorted, while the far end remained open.

The EM structure was embedded within an AWR schematic environment for parameter extraction. The left and right feedline ports were each terminated through a $50~\Omega$ resistor to schematic ports $P1$ and $P2$, respectively, while the resonator port was connected directly to schematic port $P3$. All admittance and transmission measurements were extracted from the schematic environment.

\subsubsection{Admittance-Based Resonance and Coupling Extraction}

The exported simulation data consisted of frequency-dependent complex admittance parameters together with feedline transmission data. The resonator self-admittance was obtained from \[Y_{33},\] while the transfer admittances between the resonator and the feedline ports were obtained from\[Y_{13} \quad \text{and} \quad Y_{23}.\]
Feedline transmission was additionally monitored using the simulated $S_{21}$ response.

The resonance frequency was extracted from the zero crossing of the imaginary component of the resonator self-admittance,\[\mathrm{Im}\left[Y_{33}\right] = 0,\] using linear interpolation between the two frequency points surrounding the crossing.

The effective resonator capacitance was determined from the slope of the imaginary admittance response at resonance,\[C_{\mathrm{eff}}=\left.\frac{d\,\mathrm{Im}[Y_{33}]}{d\omega}\right|_{\omega=\omega_r}.\]

The effective external conductance was calculated from the transfer admittances according to\[R_{\mathrm{eff}}=Z_0\left(|Y_{13}|^2+|Y_{23}|^2\right),\]

where

\[
Z_0 = 50~\Omega.
\]

The external coupling rate was then calculated as

\[
\kappa
=
\frac{R_{\mathrm{eff}}}
{C_{\mathrm{eff}}},
\]

and reported in MHz according to

\[
\kappa_{\mathrm{MHz}}
=
\frac{R_{\mathrm{eff}}}
{C_{\mathrm{eff}}\,2\pi \times 10^6}.
\]

The external quality factor was calculated from

\[
Q_{\mathrm{ext}}
=
\frac{f_r}{\kappa}.
\]

The resonator length, coupling gap, and coupling-section length were adjusted to achieve the desired resonance frequency and target external quality factor of approximately $Q_{\mathrm{ext}} \sim 10^5$. The three resonator design targeted resonances near 4.7, 5.2, and 5.8~GHz, with resonators alternated above and below the feedline to reduce spatial crowding and undesired resonator-resonator coupling.

\begin{table*}[htbp!]
    \centering
    \caption{
    Summary of the simulated InP 35/20 resonator design parameters. Resonators are ordered from left-to-right along the feedline. The simulated external quality factors were designed to target approximately $Q_{\mathrm{ext}} \sim 10^5$.
    }
    \label{tab:sim_summary}
    \begin{tabular}{cccccccccc}
    \hline
    Resonator & Simulated $f_r$ (GHz) & $Q_{\mathrm{ext}}$ & Gap ($\mu$m) & Coupling Length ($\mu$m) & $\kappa$ (MHz) & $C_{\mathrm{eff}}$ (fF) \\
    \hline
    A & 4.5216 & 101,902 & 77.05 & 230.00 & 0.04437 & 961.3 \\
    B & 5.4608 & 97,479 & 100.00 & 200.50 & 0.05602 & 891.6 \\
    C & 5.7383 & 102,920 & 100.00 & 164.92 & 0.05576 & 860.8 \\

    \hline
    \end{tabular}
\end{table*}

\section{Resonator Fabrication}
\label{Res_fab}
\begin{figure*}[htbp!]
    \centering
    \includegraphics[width=0.8\linewidth]{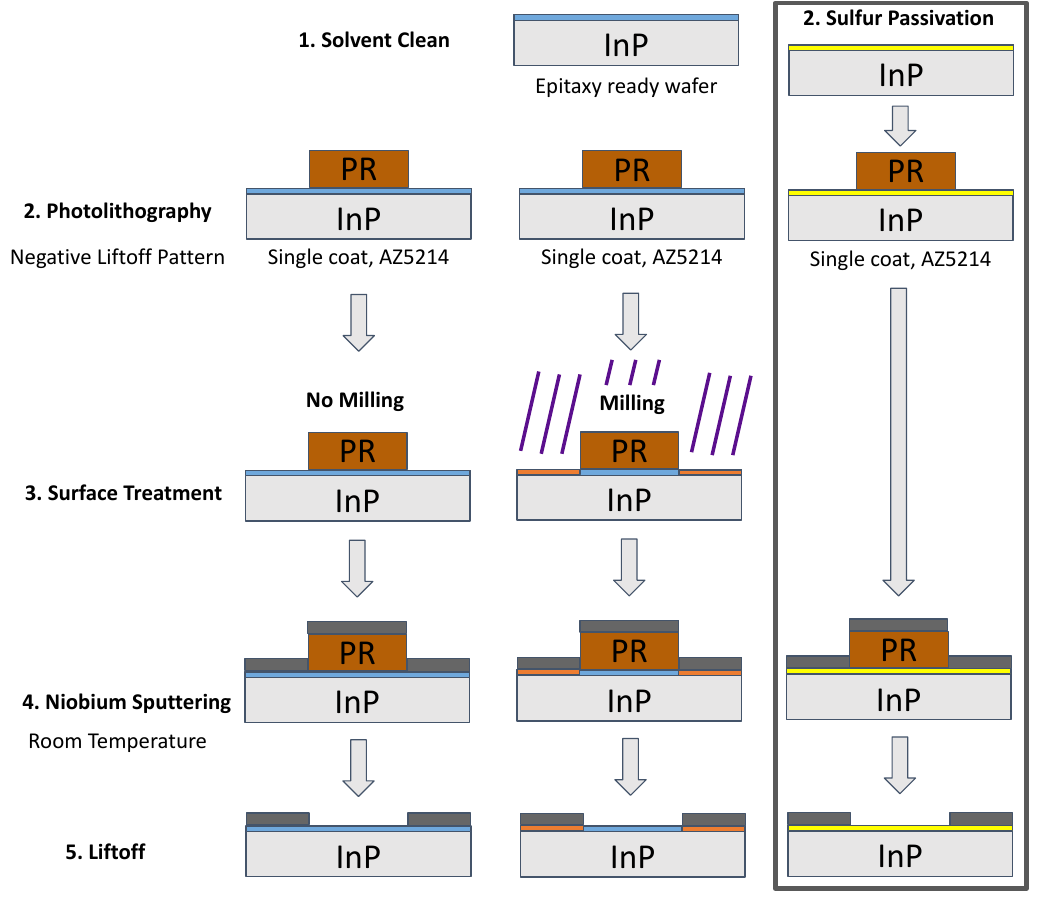}
    \caption{
    Fabrication flow for processing coplanar waveguide resonators.
    }
    \label{fig:s_process_flow}
\end{figure*}

\subsection{Substrate Preparation and Nb Deposition}

A graphical process flow is depicted in Figure~\ref{fig:s_process_flow}. Semi-insulating Fe-doped InP substrates (Acrotec, (001), $\rho = 2.3 \times 10^{7}\ \Omega\cdot\text{cm}$) were cleaved into $7 \times 7$ mm chips prior to solvent cleaning. Substrates were first solvent cleaned using a 10 minute acetone soak at $50^\circ$C, followed by an isopropyl alcohol (IPA) rinse. For lift-off patterning, chips were spin-coated with AZ-5214 photoresist and soft baked at $110^\circ$C for 1 minute prior to i-line contact lithography exposure with a designed critical dimension (CD) of $20\ \mu$m. Development was performed using a tetramethylammonium hydroxide (TMAH)-based developer to produce a lift-off profile suitable for deposition of up to 200 nm thick metal films.

To maximize compatibility with standard Josephson junction fabrication materials and processing workflows, all sputtering was performed at room temperature. Niobium films with a nominal thickness of 130 nm were deposited in a high-vacuum DC sputtering chamber with a base pressure of $5 \times 10^{-9}$ Torr. The sputtering process used 80 W DC power with 5 sccm argon flow. Prior to deposition, the plasma plume was stabilized and the Nb target was cleaned using a 10-minute pre-sputter cycle. Film deposition was then performed for 60 minutes.

\subsection{Sulfur Passivation Process}

Prior studies have shown that sulfur passivation improves the electrical characteristics of metal--oxide--semiconductor capacitors fabricated on InP \cite{lee2019facile}. Motivated by this, sulfur passivation was applied prior to Nb deposition to suppress native oxide regrowth and improve buried interface quality. To remove the native oxide, a 1\% HF buffered oxide etch (BOE) solution was prepared by diluting 10~mL of stock 6:1 BOE (Transene Company) with 70~mL of DI water. In a separate adjacent beaker, 60 mL of 10\% $(\mathrm{NH}_4)_2\mathrm{S}$ solution was prepared by diluting a 20\% aqueous ammonium sulfide stock solution (Thermo Fisher Scientific). Two large deionized water rinse baths and a third cascade rinse bath were prepared prior to processing. Depending on device type, chips were loaded either patterned or unpatterned into a PTFE dipper and submerged in the 1\% HF oxide removal bath for 30 seconds. Samples were then immediately transferred into the ammonium sulfide bath for a 10-minute soak. During this treatment, surface hydroxyl and native oxide species are replaced by sulfur, resulting in the formation of a chemically bound sulfur termination layer. High-resolution photoemission studies have shown that this treatment produces approximately one monolayer of sulfur coverage on the InP surface, with sulfur atoms forming bridge bonds exclusively to surface indium atoms \cite{Tao1992}. In this configuration, each sulfur atom bonds to two neighboring indium atoms along the surface, occupying a position analogous to the top-layer phosphorus site while remaining slightly displaced out of plane. The absence of detectable P--S bonding indicates that phosphorus is not retained in the passivated surface layer, and that the termination is instead dominated by In--S bonding. This bonding configuration effectively saturates the dangling bonds of surface indium atoms, preventing the formation of surface dimers and stabilizing a $(1\times1)$ surface structure. As a result, the sulfur-passivated surface exhibits greatly reduced surface state density and enhanced resistance to native oxide regrowth. In the presence of residual acidity, sulfide ions may react according to 

\begin{equation}
    \mathrm{S}^{2-} + 2\mathrm{H}^+ \rightarrow \mathrm{H}_2\mathrm{S}(g)
\end{equation}

resulting in the formation of small quantities of hydrogen sulfide gas. Accordingly, all processing was conducted under maximum fume hood draft to ensure safe ventilation.

Following the 10 minute sulfide soak, samples were sequentially transferred through two short dilution rinse baths, followed by an extended cascade rinse. The cascade rinse was continued until the rinse bath reached a stable pH endpoint, ensuring removal of residual acidic species while preserving the sulfur termination. Samples were loaded into the sputtering chamber load lock within 10 minutes of completion of the wet chemical treatment to minimize contamination and surface reoxidation. Chips were mounted using Cu tape as described previously.

\subsection{Sulfur-Passivated Lift-Off Process}

Sulfur-passivated resonator devices used in the $Q_i$ measurements were fabricated using the same single-layer AZ5214 lift-off lithography process as the non-passivated control devices. The sulfur passivation step was performed prior to photoresist coating, directly on the cleaned substrate surface. Therefore, the photoresist was not exposed to the large pH excursion associated with the sulfur passivation chemistry.

The initial substrate preparation remained unchanged: a 10 minute acetone soak at $50^\circ$C, IPA rinse and nitrogen dry. After this cleaning sequence, the sulfur passivation process was applied to the bare substrate. A 2-minute dehydration bake at $110^\circ$C was then done to ensure all liquid from the passivation rinses was evaporated. The passivated substrate was then coated with AZ5214 at 4000 RPM and patterned using the same exposure dose and development conditions as the other single-layer AZ5214 lift-off chips.

Consequently, the lithographic linewidth, sidewall profile, and lift-off geometry were held constant between sulfur-passivated and non-passivated resonators. Any differences in the measured $Q_i$ can therefore be attributed to the sulfur-treated surface rather than to changes in the lithography process.

\section{Resonator Measurement}
\label{Resonator_measurement}

\subsection{Hardware setup}
\label{Hardware_setup}

\begin{figure*}[htbp!]
    \centering
    \includegraphics[width=6.5in]{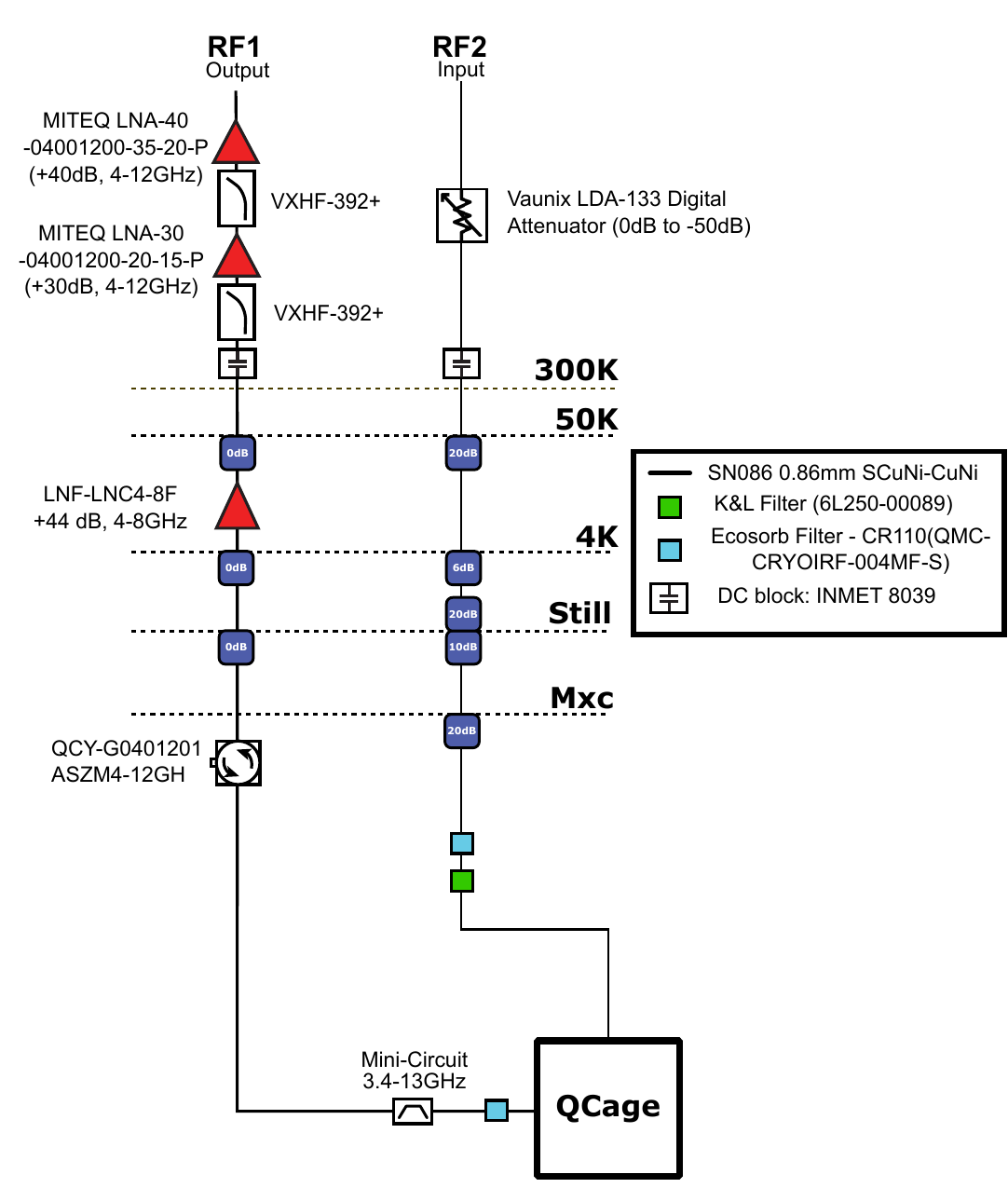}
    \caption{Wiring diagram for resonator RF measurement in a Bluefors SD system}
    \label{fig:s_measurement_setup}
\end{figure*}

CPW resonators fabricated with the three surface treatments were measured in a Bluefors SD dilution refrigerator equipped with an RF measurement setup. The resonator chip is mounted inside a Qdevil Qcage.24, an RF cavity sample holder designed to suppress internal resonance up to 18GHz, well above the operating bandwidth of the output line (4-8GHz). 

\begin{figure*}[htbp!]
    \centering
    \includegraphics[width=0.9\linewidth]{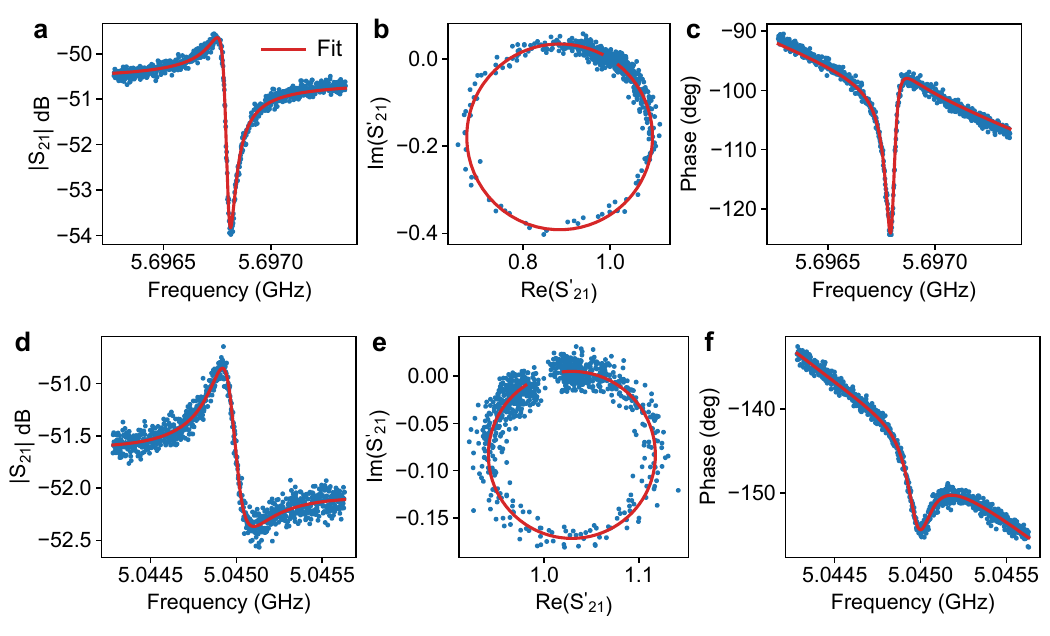}
    \caption{
    Representative resonator fits for the as-received and milled devices.
    Panels \textbf{(a--c)} show the lowest-power fit for the as-received resonator near
    $f_r = 5.70~\mathrm{GHz}$ using the complex notch-resonator circle-fitting method
    described in Section~\ref{symetric_fitting}. Panel \textbf{(a)} shows the measured
    magnitude response with the fitted model overlaid, panel \textbf{(b)} shows the
    normalized complex response in the IQ plane, and panel \textbf{(c)} shows the phase
    response.
    Panels \textbf{(d--f)} show the corresponding fit windows for the asymmetric milled
    resonator near $f_r = 5.04~\mathrm{GHz}$ at high power,
    fit using the closest-pole-and-zero method described in
    Section~\ref{asymetric_fitting}. Panel \textbf{(d)} shows the magnitude response,
    panel \textbf{(e)} shows the normalized complex response, and panel \textbf{(f)}
    shows the phase response. In each panel, blue points are measured data and the red
    curve is the fitted model.
    }
    \label{fig:milled_fit}
\end{figure*}

The measurement setup used in this study, shown in Figure~\ref{fig:s_measurement_setup}, consists of an attenuated input line and an amplified output line for $|S_{21}|$ readout. The input line starts at port 1 of a Keysight ENA E5063A Vector Network Analyzer (VNA). For the highest-power measurements, the VNA output power is swept from $+10$ to $+5$~dBm with the Lab Brick digital attenuator set to 0~dB attenuation. For lower-power measurements, the VNA output power is held fixed at $+5$~dBm while the Lab Brick digital attenuator is swept from 0 to $50$~dB, maximizing the VNA output power to maintain the highest possible signal-to-noise ratio. The VNA intermediate-frequency (IF) bandwidth is set to 5~Hz for effective source-power settings from $+10$ down to $-34$~dBm, and is reduced to 1~Hz for effective source-power settings from $-35$ to $-45$~dBm, corresponding to Lab Brick attenuation settings from $40$ to $50$~dB with the VNA fixed at $+5$~dBm.

After the Lab Brick attenuator, the input line enters the fridge column through an INMET DC block. From the 50~K stage to the mixing chamber, a total of $-76$~dB attenuation is added to the line. Before the input reaches the Qcage, it is filtered by an Eccosorb filter with a 10~GHz cutoff and a K and L bandpass filter with a 12~GHz cutoff. The total insertion loss of the remaining input-line components is determined from datasheets at the median resonator frequency, approximately 5~GHz, to be $-23$~dB.

The output signal first passes through an Eccosorb filter and a Mini-Circuits bandpass filter (3.4-13GHz) after the Qcage output. A single Quinstar isolator is used to suppress reflections from the higher-temperature output chain before the signal is routed up the fridge stages and amplified by an LNF 4~K high-electron-mobility transistor (HEMT) (+44dB) amplifier. At room temperature, an INMET DC block is connected to the top flange of the fridge. A MITEQ LNA-30 HEMT (+30dB) and a MITEQ LNA-40 amplifier (+40dB), both operating from 4 to 12~GHz, are then added to bring the signal above the noise floor of the VNA. A VXHF-392+ reflectionless high-pass filter is placed before each of the room temperature MITEQ LNAs to prevent saturation and eliminate out-of-band reflections. The output is then connected directly to port 2 of the VNA.

As the frequency is swept from 4 to 10~GHz, the transmission coefficient, $S_{21}$, is measured. Resonators appear as dips in the transmission coefficient at their respective resonance frequencies. After identifying each resonance, fine sweeps with a 3~MHz span are performed around the resonance frequencies.

\section{Fitting}
\label{fitting}
\subsection{Symmetric resonator fitting}
\label{symetric_fitting}
The measured resonator response appears as a notch in the complex transmission coefficient $S_{21}$. Although the magnitude response is often approximately Lorentzian, all fits in this work are performed using the full complex-valued $S_{21}$ data. We use the complex notch-resonator fitting procedure described by Baity et al.~\cite{Baity2024}, which accounts for electrical delay, complex background rotation, and impedance-mismatch asymmetry.

The fitted response is \[S_{21}(f) = a e^{i\alpha - 2\pi i f \tau} \left[1 - \frac{Q_l}{|Q_c|}\frac{e^{i\phi}}{1 + 2 i Q_l \left(f/f_r - 1\right)} \right].\] Here $f_r$ is the resonance frequency, $Q_l$ is the loaded quality factor, $|Q_c|$ is the magnitude of the coupling quality factor, $\phi$ is the asymmetry angle, $\tau$ is the electrical delay, and $a e^{i\alpha}$ is an overall complex prefactor.

Following Baity et al.~\cite{Baity2024}, the diameter-corrected coupling
quality factor is
\[
Q_c^\mathrm{dia} = \frac{|Q_c|}{\cos\phi},
\]
and the corresponding internal quality factor is
\[
\frac{1}{Q_i^\mathrm{dia}}
=
\frac{1}{Q_l}
-
\frac{1}{Q_c^\mathrm{dia}}.
\]
We also record the uncorrected value
\[
\frac{1}{Q_i^\mathrm{no\,corr}}
=
\frac{1}{Q_l}
-
\frac{1}{|Q_c|}.
\]

The on-chip photon occupation is calculated using the expression from
Baity et al.~\cite{Baity2024},
\[
\bar{n}
=
\frac{P_\mathrm{chip}}{2\pi h f_r^2}
\left(
\frac{2 Q_l^2 \cos\phi}{|Q_c|}
\right).
\]
Here $P_\mathrm{chip}$ is the microwave power at the device in watts. In this
experiment it is computed from the VNA output power, the Lab Brick digital
attenuation, and the fixed fridge and component attenuation:
\[
P_\mathrm{chip}[\mathrm{dBm}]
=
P_\mathrm{VNA}[\mathrm{dBm}]
-
A_\mathrm{dig}[\mathrm{dB}]
-
99~\mathrm{dB}.
\]

Uncertainties are estimated using the residual-based covariance procedure
described by Baity et al.~\cite{Baity2024}. The reported uncertainties in
$Q_l$, $|Q_c|$, $f_r$, and $\phi$ are propagated to obtain the uncertainties in
$Q_i^\mathrm{dia}$ and $Q_i^\mathrm{no\,corr}$. 

\subsection{Asymmetric resonator fitting}
\label{asymetric_fitting}

For most resonances, the complex notch response is fit using the diameter-corrected circle-fit method described above. The milled resonator near $5.04~\mathrm{GHz}$, however, exhibited a strongly asymmetric line shape, consistent with parasitic reflections arising from impedance discontinuities in the device or microwave measurement chain~\cite{Deng2013}. The $\sim 5.04~\mathrm{GHz}$ resonance on the milled chip was therefore treated separately. For this resonance, the Baity circle-fit model returned an anomalously large coupling quality factor, $|Q_c| > 10^7$, and consequently photon occupations that were orders of magnitude lower than those obtained for neighboring resonators measured under the same power calibration. We therefore do not use the Baity-derived $Q_c$, $Q_i$, or photon number for this resonance.

Instead, the milled $\sim 5.04~\mathrm{GHz}$ resonance is fit independently
using the Closest Pole and Zero Method (CPZM) of Deng, Otto, and
Lupascu~\cite{Deng2013}. In this approach, the local complex transmission is
modeled as a rational pole-zero response with a smooth complex background,
\[
S_{21}(f)
=
B(f)
\frac{x-z}{x-p},
\qquad
x=\frac{f-f_\mathrm{ref}}{\Delta f}.
\]
Here $B(f)$ is a complex linear background, $p=p_r+i p_i$ is the fitted pole,
and $z=z_r+i z_i$ is the fitted zero. This model captures the asymmetric
line shape through the pole-zero geometry rather than through the Baity
diameter-correction angle $\phi$.

The resonance frequency and loaded quality factor are extracted from the pole:
\[
f_r = f_\mathrm{ref} + p_r \Delta f,
\qquad
\gamma = |p_i|\Delta f,
\qquad
Q_l = \frac{f_r}{2\gamma}.
\]
The coupling is extracted from the pole-zero separation,
\[
d_\mathrm{pz}
=
\frac{p-z}{i p_i},
\qquad
|Q_c|_\mathrm{pz}
=
\frac{Q_l}{|d_\mathrm{pz}|},
\]
and the internal quality factor is calculated as
\[
\frac{1}{Q_i^\mathrm{pz}}
=
\frac{1}{Q_l}
-
\frac{1}{|Q_c|_\mathrm{pz}}.
\]

Because this method does not use the Baity asymmetry angle, the exported values
for this resonance are reported with $\phi=0$ and
\[
Q_c^\mathrm{dia}=|Q_c|_\mathrm{pz},
\qquad
Q_i^\mathrm{dia}=Q_i^\mathrm{no\,corr}=Q_i^\mathrm{pz}.
\]
The same calibrated on-chip power is used as for the other resonators:
\[
P_\mathrm{chip}[\mathrm{dBm}]
=
P_\mathrm{VNA}[\mathrm{dBm}]
-
A_\mathrm{dig}[\mathrm{dB}]
-
99~\mathrm{dB}.
\]
The photon occupation for this resonance is calculated using the pole-zero
coupling quality factor,
\[
\bar{n}_\mathrm{pz}
=
\frac{P_\mathrm{chip}}{2\pi h f_r^2}
\left(
\frac{2 Q_l^2}{|Q_c|_\mathrm{pz}}
\right).
\]
Thus, the power calibration is unchanged, but the coupling quality factor used
to convert power to photon number is obtained from the CPZM pole-zero fit
rather than from the failed Baity circle fit.

\bibliographystyle{apsrev4-2}
\bibliography{ABIB}

\end{document}